\RequirePackage{lineno}
\documentclass[pdflatex,aps,prd,twocolumn,nofootinbib,showpacs,superscriptaddress]{revtex4-2}
\usepackage{amsmath,amssymb,color,fancyhdr,hyperref}
\usepackage{graphicx}
\usepackage[latin1]{inputenc}
\usepackage[english]{babel}
\usepackage{booktabs, multirow, makecell}%
\usepackage{comment}
\usepackage{capt-of}
\usepackage{float}

\begin{document}
\title{Scaling functions of the three-dimensional \boldmath $Z(2)$, $O(2)$, and $O(4)$ models and their finite size dependence in an external field}

\author{F.~Karsch}
\affiliation{Fakult\"at f\"ur Physik, Universit\"at Bielefeld, D-33615 Bielefeld, Germany}
\author{M.~Neumann}
\affiliation{Fakult\"at f\"ur Physik, Universit\"at Bielefeld, D-33615 Bielefeld, Germany}
\author{Mugdha Sarkar}
\affiliation{Physics Division, National Center for Theoretical Sciences, National Taiwan University, Taipei 10617, Taiwan}
\begin{abstract}
We analyze scaling functions in the $3$-$d$, $Z(2)$, $O(2)$ and $O(4)$ 
universality classes and their finite size dependence using Monte Carlo simulations of improved $\phi^4$ models.
Results for the scaling functions are fitted to the 
Widom-Griffiths form, using a parametrization also used in analytic calculations.
We find good agreement on the level
of scaling functions and the location
of maxima in the universal part of susceptibilities. We also find that an earlier parametrization of the $O(4)$ scaling function, using 14 parameters, is well reproduced when using the Widom-Griffiths form with only three parameters.
We furthermore show that finite size corrections to the scaling functions are distinctively different in the $Z(2)$ and $O(N)$ universality classes and determine the volume dependence of the peak locations in order parameter and mixed susceptibilities.
\vspace{0.2in}
\begin{center}
\bf{\today}
\end{center}
\end{abstract}

\vspace{0.2in}

\newcommand \tb {\bar{t}}

\maketitle

\section{Introduction}

Universal critical behavior in the 
$3$-$d$, $Z(2)$ and $O(N)$ universality classes plays an important role in the analysis of phase transitions in many statistical models as well as quantum field theories.
In the fundamental theory of strong
interactions, Quantum Chromodynamics (QCD), phase transitions that occur at finite temperature and vanishing as well as non-vanishing conserved charge chemical potentials belong to these 
universality classes.
The spontaneous breaking of chiral
symmetry in QCD is expected to exhibit
universal critical behavior in the 
$3$-$d$, $O(4)$ universality class
\cite{Pisarski:1983ms}, and
the $3$-$d$, $Z(2)$ universality class 
is expected to describe critical behavior at the
so-called critical endpoint,
a yet to be discovered second order phase transition that is expected to occur in QCD with non-zero quark mass 
values and non-vanishing baryon chemical potential. A second order 
phase transition in the $Z(2)$
universality class also occurs in QCD at non-vanishing, imaginary values of chemical potentials at the so-called 
Roberge-Weiss endpoint \cite{Roberge:1986mm}. 
Also, the $O(2)$ universality class plays 
a role in the studies of the phase diagram of QCD, as many calculations are 
performed in a discretized version of the theory, using so-called staggered fermions, in which only this smaller
symmetry is realized.

Numerical studies of the phase structure of statistical 
models, and, in particular, of complicated theories such as QCD, 
are being performed on finite
lattices. A good understanding of finite-size effects,
thus, is generally of importance.
In the limit of small external symmetry breaking fields 
and large volumes also these finite-size
effects are universal, {\it i.e.}, characteristic for a given universality
class. For this purpose a powerful
renormalization group framework has been developed in statistical physics 
which leads to a detailed finite-size scaling 
theory for critical behavior \cite{Fisher:1972zza,Privman:1984zz}.
This framework has been used to analyze finite-size scaling
behavior of systems in the $3$-$d$, $Z(2)$ and $O(N)$ 
universality classes. The finite-size dependence of thermodynamic 
observables in $3$-$d$, $O(N)$ spin models has 
been examined using Monte Carlo simulations 
\cite{Engels:1999wf,Engels:2000xw,Campostrini:2000iw}, and 
finite-size scaling functions have been derived using the 
functional renormalization group approach \cite{Springer:2015kxa}.
For the $O(4)$ universality class, we provided
an updated parametrization for the infinite volume scaling functions \cite{Engels:2011km}
and presented a parametrization of the finite-size scaling functions
$f_G(z,z_L)$ and $f_\chi(z,z_L)$
\cite{Engels:2014bra},
which describe finite volume corrections to the singular behavior of the order parameter 
and its susceptibility. In this 
work, we will extend these studies
and provide finite-size scaling
functions also for the $Z(2)$
and $O(2)$ universality classes, by performing 
Monte Carlo simulations with improved Hamiltonians
\cite{Ballesteros:1998my,Hasenbusch:1999mw,Hasenbusch:1999cc},
which have been constructed to suppress contributions
from corrections-to-scaling and, thus, allow for
easier access to the desired scaling functions.
We furthermore present a parametrization of the infinite volume
scaling functions, determined from Monte Carlo simulations, 
using the Widom-Griffiths (WG) form
\cite{Widom,PhysRev.158.176,PhysRevLett.22.606,PhysRevLett.23.1098} of these
scaling functions. In the 
$Z(2)$ \cite{Zinn-Justin:1999opn} and $O(2)$ \cite{Campostrini:2000iw}
universality classes, the relevant parameters entering this
analytic form have been determined previously using $\epsilon$-expansion and other
field theoretic methods applied directly in $3$-$d$. 

This paper is organized as follows. In the next section we introduce the $Z(2)$ and $O(N)$ models for which we present new Monte Carlo results and define the basic observables studied by us. In Section III we introduce basic relations for finite-size scaling functions. Section IV is devoted to the determination of the infinite volume scaling functions
for the $Z(2)$, $O(2)$ and $O(4)$ models,
using a parametrization based on 
the Widom-Griffiths form.
Here, we also determine the non-universal parameters for 
the improved $Z(2)$ and $O(2)$ models
that are needed to introduce the 
scaling variables $z$ and $z_L$.
In Section V we present our results for the finite-size scaling functions. We give our conclusions in Section VI. In  
Appendix \ref{app:H0L0}, we discuss the determination of the two non-universal scales $H_0$ and $L_0$ and in Appendix \ref{app:Widom}, we give explicit expressions for the expansion 
coefficients $d_1^-$ and $d_2^-$ appearing in the scaling function $f_G(z,z_L=0)$ at asymptotically large, negative arguments.

\section{Lattice setup and observables}
\label{sec:setup}
We discuss here universal scaling properties for  $3$-dimensional, $N$-component $\phi^4$ models, 
{\it i.e.},
spin models in the $3$-$d$, $Z(2)$ ($N=1$, Ising model), $O(2)$ (XY model) and
$O(4)$ universality classes described by the
Hamiltonian,
\begin{eqnarray}
 \mathcal{H} &=& -J \sum_{\left\langle x,y \right\rangle} \Phi_x \Phi_y + \sum_x \left[ \Phi^2_x + \lambda \left(  \Phi^2_x -1 \right)^2 \right]\nonumber \\
    &&- H \sum_x \phi_{x,1} \; ,
    \label{Hamiltonian}
\end{eqnarray}
with $\Phi_x\equiv \phi_{x,1}$ for the 3-$d$, $Z(2)$ spin model, $(x,y)$ 
denoting nearest neighbor sites on the
lattice, and $\Phi_x\equiv (\phi_{x,1},...,\phi_{x,N})$ for the 3-$d$, $O(N)$ spin models.
For specific choices of $\lambda$,
the above Hamiltonian is called ``improved" since the quartic coupling $\lambda$ appearing in the potential term of the spin models has been optimized to reduce the effect of contributions from sub-leading relevant scaling variables to universal scaling behavior of these models \cite{Ballesteros:1998my}.
We use the parameters $\lambda=1.1$ \cite{Hasenbusch:1999mw} 
in the case of $Z(2)$ and $\lambda=2.1$ for $O(2)$ spin models 
\cite{Hasenbusch:1999cc}, respectively. In the $O(4)$ case,
we use the standard, unimproved Hamiltonian, corresponding to 
$\lambda=\infty$.
The temperature $T$ is defined as the inverse of the coupling $J$, {\it i.e.}, $T\equiv 1/J$, and the external field coupling $H$ controls explicit symmetry 
breaking in the Hamiltonian.
We introduce this symmetry breaking term such that it
couples only to the first component of the spin variable $\Phi_x$, defined on the sites $x$ of a three dimensional lattice of size $L^3$.

Using the Hamiltonian introduced in Eq.~\ref{Hamiltonian} the partition functions of the $3$-$d$,
$Z(2)$ and $O(N)$ models are given by,
\begin{equation}
  Z(T,H,L) =\int \prod_x {\rm d}\Phi_x {\rm e}^{-\mathcal{H}} \; .
  \label{partition}
\end{equation}
From this, one obtains the free energy density in units of temperature $T$, $f(T,H,L)=-L^{-3}\ln Z(T,H,L)$.
The derivative of the free energy density with respect to the external field $H$ defines the order parameter, $M$, for spontaneous symmetry breaking, 
\begin{equation}
 M (T,H,L) = - \frac{\partial f}{\partial H} \;\; . \;\;
 \label{M}
\end{equation}
The (longitudinal) susceptibility $\chi_h$ and the 
mixed susceptibility $\chi_t$ are 
obtained as derivatives of the order 
parameter with respect to $H$ and $J$, respectively,
\begin{eqnarray}
 \chi_h(T,H,L) &=&  
 \frac{\partial M}{\partial H} \; ,
 \label{chih} \\
    \chi_t(T,H,L) &=&\frac{
    \partial M}{\partial J} = -
    T^2\frac{\partial M}{\partial T} \;\; .
    \label{chit}
\end{eqnarray}

\begin{table}[t]
\centering
\begin{tabular}{|c|c|c|c|} \hline
 & $Z(2)$ & $O(2)$ & $O(4)$ \\ \hline
  \multicolumn{4}{|c|}{universal parameter}\\ \hline
  $\beta$ & 0.3258(14) & 0.34864(7) & 0.380(2) \\
$\delta$ & 4.805(15) & 4.7798(5) & 4.824(9) \\ \hline
  \multicolumn{4}{|c|}{non-universal parameter}\\ \hline
$\lambda$ & 1.1 & 2.1 & $\infty$ \\ \hline
$T_c$ & ~2.665980(3)~ & ~1.964055(23)~& ~1.06849(11)~~\\
$L_0$ & 1.0262(18) & 0.97917(55) &  0.7686\\ 
$H_0$ & 0.79522(17) & 1.36632(28) &  4.845(66)\\
$t_0$ & 0.303376(45) & 0.4540(11) & 1.023(16)\\
\hline
\end{tabular}
\caption{Critical exponents in the
$3$-$d$, $Z(2)$, $O(2)$ and $O(4)$ universality classes and non-universal parameters
for the improved Hamiltonians used 
in our simulations.
$Z(2)$ critical exponents are taken from Zinn-Justin \cite{Zinn-Justin:1999opn} and 
the $O(2)$ values are taken from \cite{Hasenbusch:2019jkj}. Exponents used for
the $O(4)$ case are taken from \cite{Engels:2003nq}.
Other critical 
exponents are obtained using the 
hyper-scaling relations. 
The critical temperature ($T_c=1/J_c$) of
the $Z(2)$ model with $\lambda=1.1$
is taken from \cite{Hasenbusch:1999mw}
and $T_c$ for the $O(2)$ model with
$\lambda=2.1$ is taken from
\cite{Campostrini:2000iw}.
All other non-universal parameters 
have been obtained in this study. For the $Z(2)$ model, we find results
for the scales $t_0$ and $H_0$ that
are in good agreement with previous results obtained in \cite{Engels:2002fi}; $t_0$ agrees to better than 1\%, and the result for $H_0$ is smaller by about 2\%.
In the $O(4)$ case, we give critical exponents and non-universal parameters used also in a previous
analysis of scaling functions \cite{Engels:2011km} (see the text for further references). 
}
\label{tab:parameter}
\end{table}

In the absence of explicit 
symmetry breaking ($H=0$), the $Z(2)$ and $O(N)$
spin models undergo second 
order phase transitions at 
critical temperatures $T_c\equiv 1/J_c$.
The critical temperatures 
of the $3$-$d$  improved $Z(2)$  \cite{Hasenbusch:1999mw}, $O(2)$
\cite{Campostrini:2000iw} and unimproved $O(4)$ \cite{Engels:2003nq} spin models, with couplings $\lambda$, as introduced above, are well determined. We give the critical temperatures together with other universal and non-universal model parameters in Table~\ref{tab:parameter}. 

In the 
case of the $O(4)$ spin model, we
do not perform new MC calculations,
but re-parametrize results for 
scaling functions already obtained in \cite{Engels:2011km}. We therefore give in Table~\ref{tab:parameter} the parameters actually used in that
calculation. They are 
consistent with analytic results
\cite{Zinn-Justin:1999opn} but 
differ somewhat from recent MC results \cite{Hasenbusch:2021rse}.

For $H\ne 0$ as well as for 
finite lattice sizes $L<\infty$,
pseudo-critical temperatures,
$T_{pc,o}(H,L)$, with $o=h$ or $t$, 
can be defined as
locations of maxima in the
susceptibilities $\chi_h$ and 
$\chi_t$. 

Monte Carlo simulations have been performed by us for the improved
$Z(2)$ and $O(2)$ models. For our
calculations we use a code, which
has been developed and used previously in simulations of $Z(2)$ and $O(2)$
models\footnote{We use a cluster update \cite{Wolff:1988uh,Wolff:1988kw} code developed in the group of
J\"urgen Engels. The algorithm and its implementation are described in
more detail in \cite{Engels:2002fi}.}.
The statistics collected in calculations with the
$Z(2)$ and $O(2)$ models on different size lattices 
is given in
Tables~\ref{tab:statistics}-\ref{tab:statistics_T0}.

\begin{table}[htb]
    \centering
    \begin{tabular}{|c|c|c|c|} \hline
      ~& ~$L=48$~ & ~$L=96$~ &~$L=120$~ \\ \hline
~$Z(2)$~~& 200000 & 100000 & - \\
~$O(2)$~~& 200000 & 200000 & 24000\\ 
\hline
    \end{tabular}
    \caption{Number of configurations generated per parameter set $(J,H)$ on lattices of size $L^3$
    for $J\ne J_c$.
    Data generated on the largest ($L=120$) lattices were used for consistency checks but were not used in the final fits.}
    \label{tab:statistics}
\end{table}

\begin{table}[htb]
    \centering
    \begin{tabular}{|c|c|c|c|} \hline
      ~& ~$L=48$~ & ~$L=96$~ &~$L=120$~ \\ \hline
~$Z(2)$~~& 200000 & 350000 & 150000 \\
~$O(2)$~~& 200000 & 100000 & 84000\\ 
\hline
    \end{tabular}
    \caption{Number of configurations generated per parameter set $(J_c,H)$, {\it i.e.} at the infinite volume critical temperature, on lattices of size
    $L^3$. These data sets were used
    for the determination of the scale parameters $H_0$ and $L_0$,
    discussed in Appendix~\ref{app:H0L0}. The data
    sets generated on the $L=96$ lattice were also used in finite-size fits discussed in Section~\ref{app:results}.}
    \label{tab:statistics_H0}
\end{table}

\begin{table}[htb]
    \centering
    \begin{tabular}{|c|c|c|c|c|} \hline
      ~& ~$L=96$~ & ~$L=120$~ &~$L=160$~  &~$L=200$~ \\ \hline
~$Z(2)$~~& 76000 & 38000 & 18000 & - \\
~$O(2)$~~& - & 480000 & 184000 & 120000 \\ 
\hline
    \end{tabular}
    \caption{Number of configurations generated per parameter set $(J,H)$ on lattices of size $L^3$ in the region $z<-2$. These data have been primarily used 
    for the determination of the scale parameter $t_0$ obtained together with the determination of the infinite volume scaling functions discussed in Section~\ref{sec:scalingfct}.}
    \label{tab:statistics_T0}
\end{table}

\section{Scaling functions}

In order to analyze universal critical behavior in the vicinity of the second order phase transitions occurring in the $3$-$d$, $Z(2)$ and $O(N)$ spin models, the free energy is split in a singular and regular contribution, respectively,
\begin{equation}
    f(T,H,L) = f_s(T,H,L) +f_{reg}(T,H,L) \; .
    \label{free}
\end{equation}
The scaling behavior of, {\it e.g.}, the order parameter $M$ and the susceptibilities $\chi_h$ and $\chi_t$ is derived from the renormalization group analysis of the singular part of the free energy
\begin{equation}
    f_s(t,h,l,\dots)= b^{-d}f_s(b^{y_t}t,b^{y_h} h,b\,l,\dots) \ ,
\label{RGfs}
\end{equation}
where $b$ is a free scale parameter and $y_t$ and $y_h$ are two relevant critical exponents\footnote{We ignore here possible contributions from corrections-to-scaling terms and irrelevant scaling fields. The former are suppressed 
in our analysis due to the use of an optimized Hamiltonian and the latter are irrelevant for the scaling analysis.}.
In Eq.~\ref{RGfs}, we introduced the reduced temperature ($t$), external field ($h$) and finite volume ($l$) scaling variables,
\begin{eqnarray}
 t =\frac{1}{t_0} \frac{T-T_c}{T_c} \; , \;
 h = \frac{H}{H_0}\; , \;
 l = \frac{L_0}{L} \; .
 \label{scalingfields}
\end{eqnarray}
They are normalized by non-universal scale parameters $t_0$, $H_0$ and $L_0$, respectively.
The exponents $y_t$ and $y_h$ define the two independent critical exponents of the universality class under consideration,
\begin{equation}
    y_t\;=\;1/\nu\;\; ,\;\;  y_h\;=\;\beta\delta/\nu  \; .
    \label{ytyh}
\end{equation}
Here $\beta,\ \delta$ and $\nu$ are critical exponents which are related to each other 
through the hyper-scaling relation $\delta = d\nu/\beta-1$.
In our current analysis we use 
results for the exponents $\beta$
and $\delta$ as basic input. 
These critical exponents are well 
determined for the $3$-$d$, $Z(2)$
and $O(N)$ universality classes.
We use here the $Z(2)$ results obtained in
\cite{Zinn-Justin:1999opn} 
and the $O(2)$ values from \cite{Hasenbusch:2019jkj}.
They are given in
Table~\ref{tab:parameter}. 
In the $O(4)$ case, we use critical exponents and non-universal parameters that have also been used
in a previous analysis of scaling functions \cite{Engels:2011km}.

Choosing the scale parameter $b=h^{-1/y_h}$, we obtain for the free energy density
\begin{equation}
    f(T,H,L)= H_0 h^{1+1/\delta} f_f(z,z_L) +f_{reg}(T,H,L)
    \; ,
    \label{free2}
\end{equation}
where we have introduced the finite-size scaling function $f_f(z,z_L)$,
\begin{equation}
    f_f(z,z_L) = H_0^{-1} f_s(t/h^{1/\beta\delta},1,l/h^{\nu/\beta\delta})\; , 
    \label{ff}
\end{equation}
with arguments $(z, z_L)$ defined as
\begin{equation}
    z = t/h^{1/\beta\delta}  \;\, , \;\; z_L = l/ h^{\nu/\beta\delta} \; .
    \label{zTzL}
\end{equation}
Using Eqs.~\ref{M} and \ref{free2} we obtain the order parameter  $M$, 
\begin{eqnarray}
 M(T,H,L) &=& h^{1/\delta} f_G(z,z_L) + reg. \;\;
 \label{Msig}
  \end{eqnarray}
and its susceptibilities
 \begin{eqnarray}
 \chi_h(T,H,L) &=& H_0^{-1} h^{1/\delta-1} f_\chi(z,z_L) + reg.
 \label{chihsig} \\
 \chi_t(T,H,L) &=& -\frac{T^2}{t_0 T_c} h^{(\beta-1)/\beta\delta} f'_G(z,z_L) +reg.
 \label{chitsig}
\end{eqnarray}
with scaling functions $f_G(z,z_L)$,
$f'_G(z,z_L)$, and $f_\chi(z,z_L)$ defined, respectively, as
\begin{eqnarray}
 f_G(z,z_L) &=&
-\left(1+\frac{1}{\delta}\right) f_f(z,z_L)
+\frac{z}{\beta\delta}\frac{\partial f_f(z,z_L)}{\partial z} \nonumber \\
&&+\frac{\nu}{\beta\delta} z_L\frac{\partial f_f(z,z_L)}{\partial z_L} \; ,
\label{fgff} \\
f'_G(z,z_L) &=& \frac{\partial f_G(z,z_L)}{\partial z} \; , 
\label{fGpff} \\
f_\chi(z,z_L) &=&
\frac{1}{\delta} \left( f_G(z,z_L)
- \frac{z}{\beta} f'_G(z,z_L) \right) \nonumber \\
&&-\frac{\nu}{\beta\delta} z_L \frac{\partial f_G (z,z_L)}{\partial z_L} \; .
\label{fchiff} 
\end{eqnarray}

The finite-size scaling functions can be determined in the vicinity of the critical point $(t,h,l)=(0,0,0)$, where regular contributions to the order parameter and its susceptibilities, given in Eqs.~\ref{Msig}$-$\ref{chitsig}, are negligible\footnote{To arrive at Eqs.~\ref{fGzzL}-\ref{fchizzL} one actually takes the limit ($H\rightarrow 0,L\rightarrow \infty$) at fixed $z_L$.},
\begin{eqnarray}
f_G(z,z_L) &=& h^{-1/\delta} M(T,H,L) \; , \label{fGzzL} \\
f'_G(z,z_L) &=& - \frac{t_0 T_c}{T^2} h^{(1-\beta)/\beta\delta}\chi_t(T,H,L)
\; , 
\label{fGprzzL} \\
f_\chi(z,z_L) &=&  H_0 h^{1-1/\delta} \chi_h(T,H,L) \; .
\label{fchizzL} 
\end{eqnarray}

The non-universal scale parameters $t_0$ and $H_0$ are fixed by the following conditions on the order parameter at infinite volume
\begin{eqnarray}
    M(t=0,h,l=0)&=&h^{1/\delta}\; , 
    \nonumber \\
    M(t<0,h=0,l=0)&=&(-t)^\beta\,,
    \label{Mconditions}
\end{eqnarray}
or, equivalently, in terms of the scaling function,
\begin{equation}
    f_G(0,0)=1 \;\; , \;\; \lim_{z\rightarrow -\infty} 
	\frac{f_G(z,0)}{(-z)^\beta}=1
    \; .
    \label{fGconditions}
\end{equation}
The scale $L_0$ is obtained using a normalization condition
for the finite-size scaling function $f_G(0,z_L)$. We define $z_L=1$ as the
point at which the order parameter, evaluated at $T_c$, is 30\% smaller than its infinite volume value, {\it i.e.}
\begin{equation}
\frac{f_G(0,1)}{f_G(0,0)}= 0.7 \; .
    \label{L01}
\end{equation}
This differs from the choice used in \cite{Engels:2014bra} but has the 
advantage of allowing better comparison of finite-size scaling functions obtained in different universality classes.

\begin{figure*}[htb]
\includegraphics[width=0.44\linewidth]{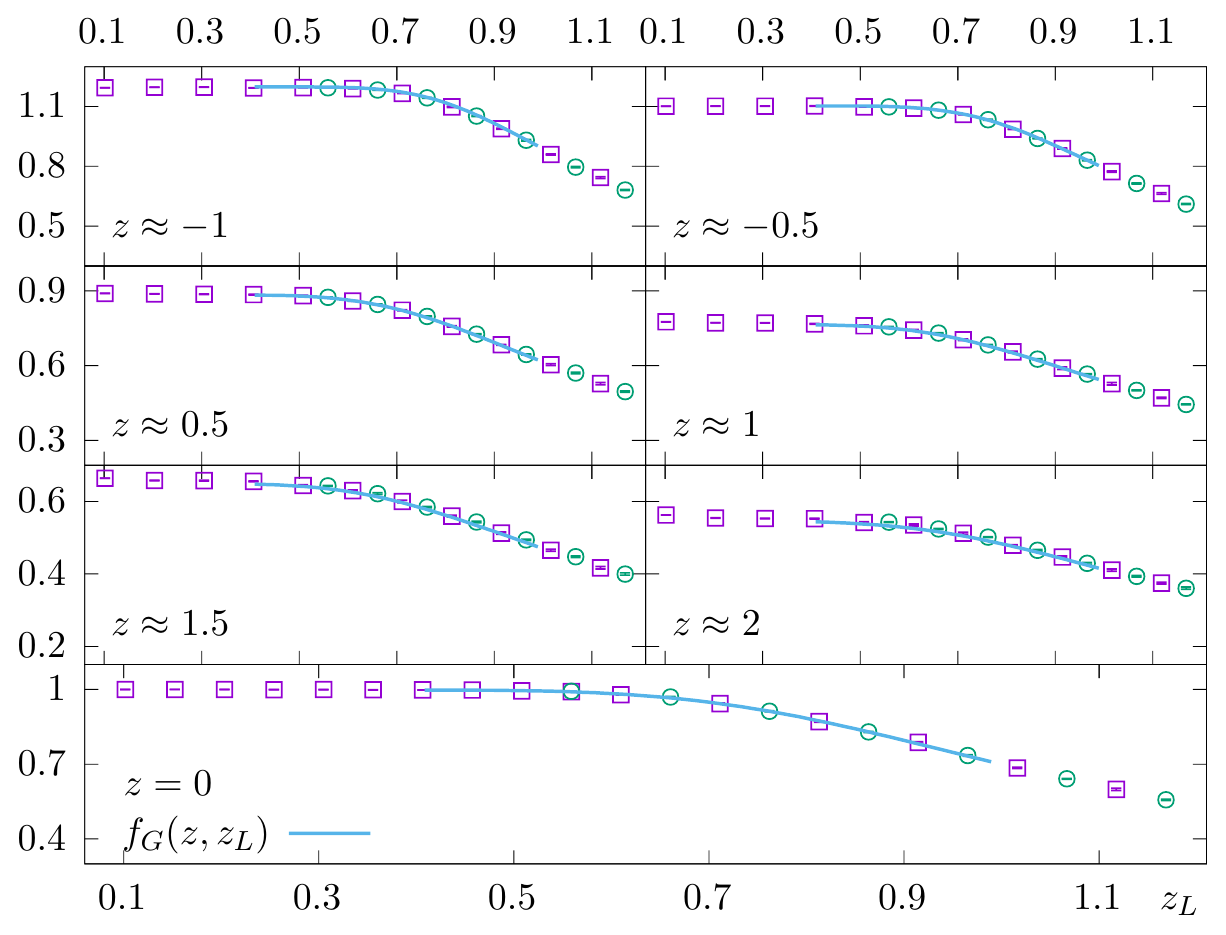}\hspace{0.7cm}
\includegraphics[width=0.44\linewidth]{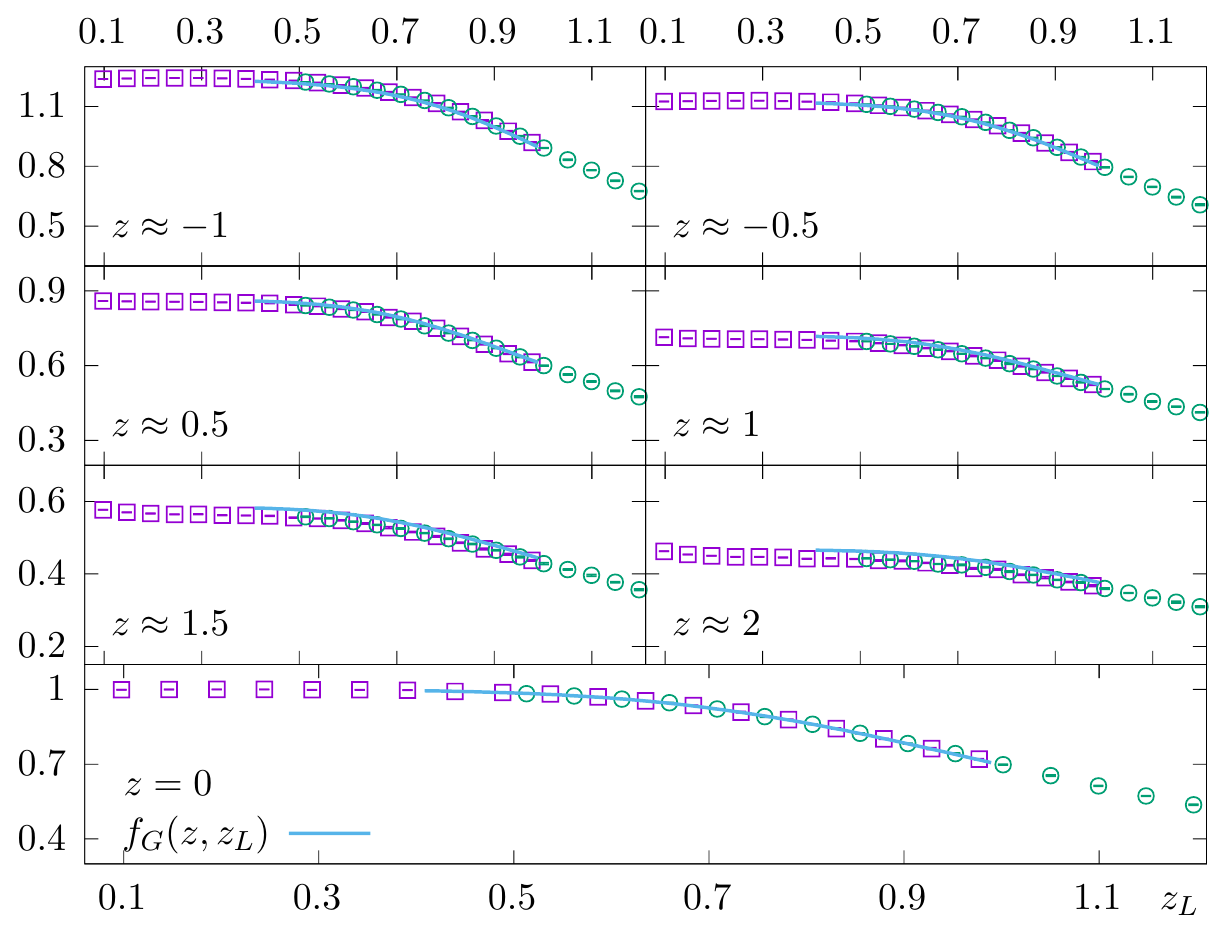}
\includegraphics[width=0.44\linewidth]{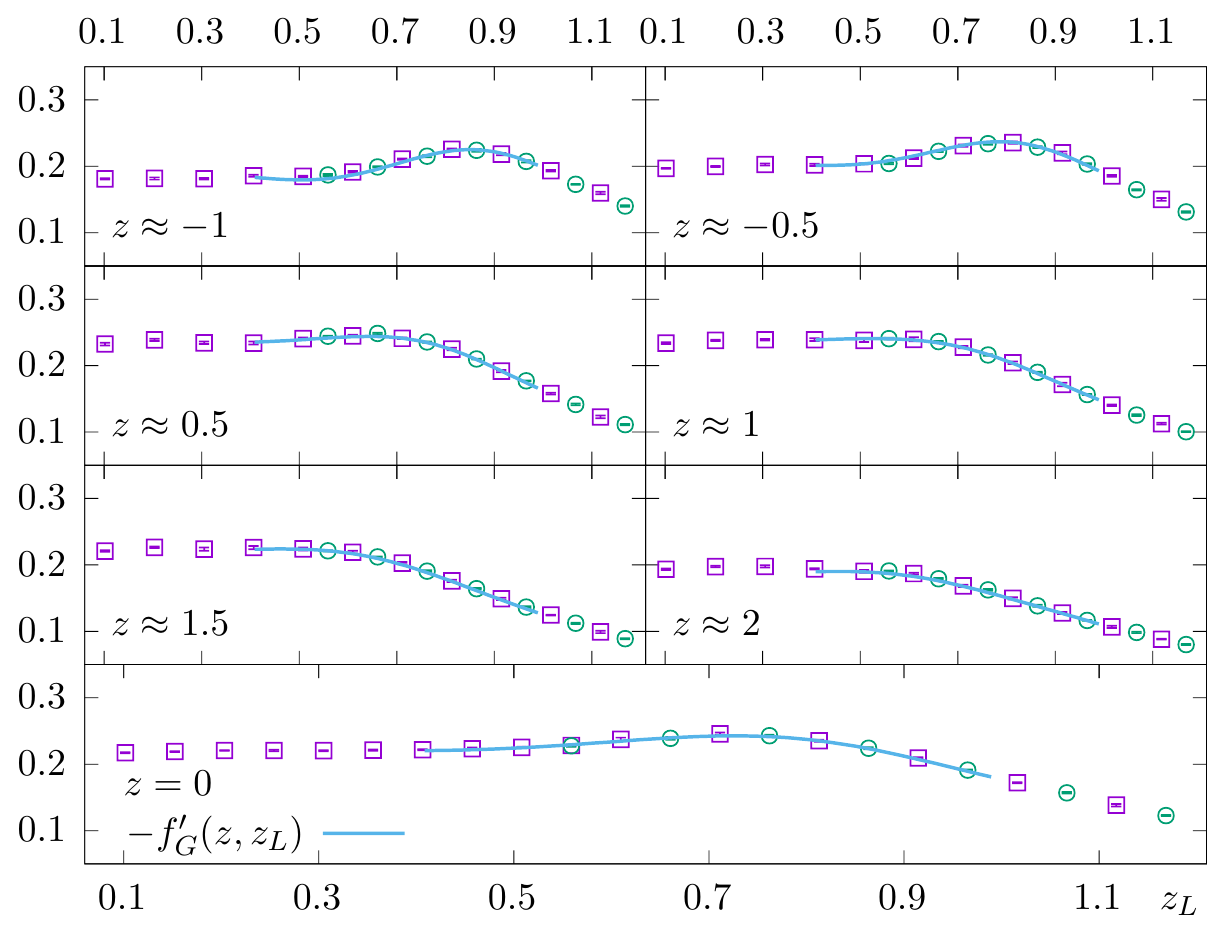}\hspace{0.7cm}
\includegraphics[width=0.44\linewidth]{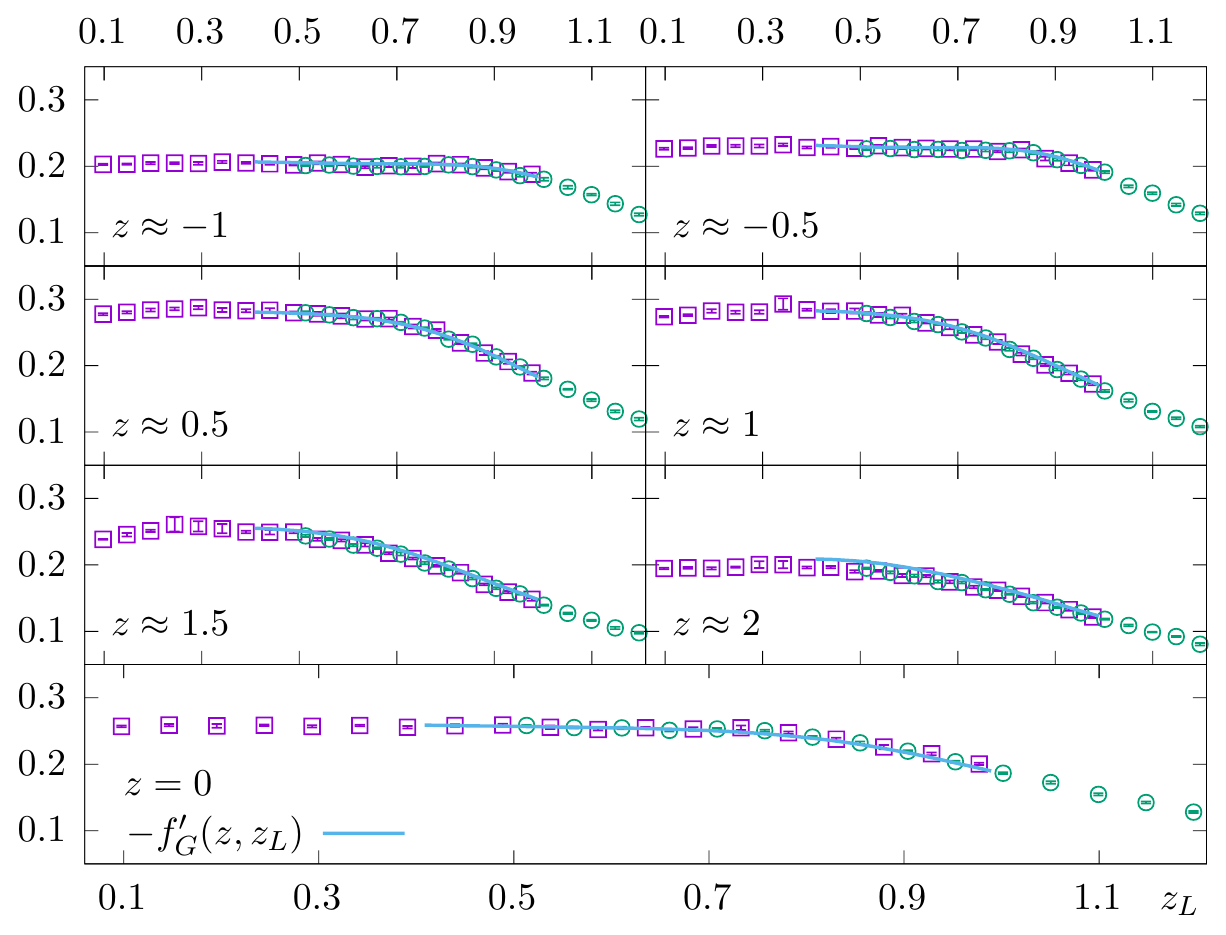}
\includegraphics[width=0.44\linewidth]{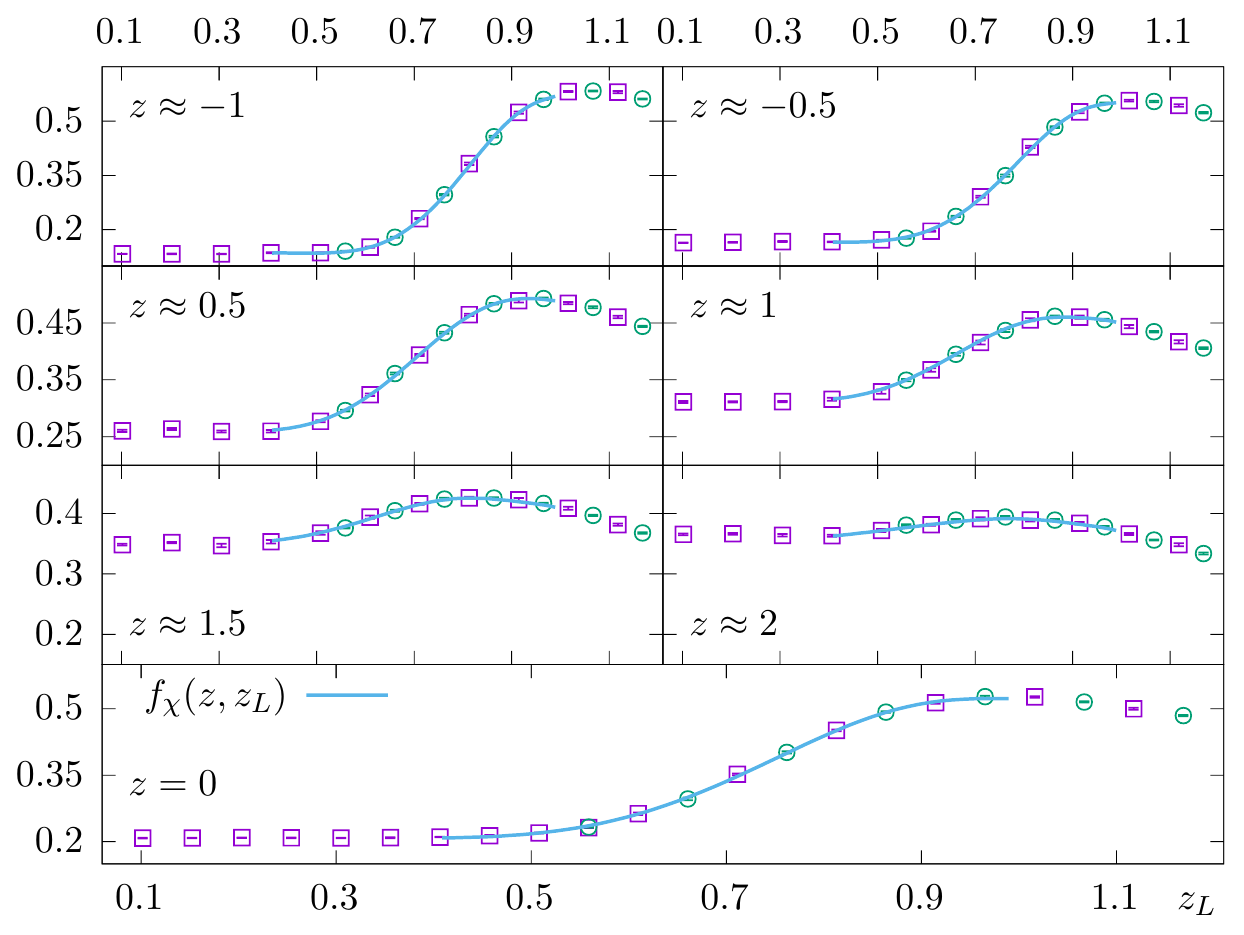}\hspace{0.7cm}
\includegraphics[width=0.44\linewidth]{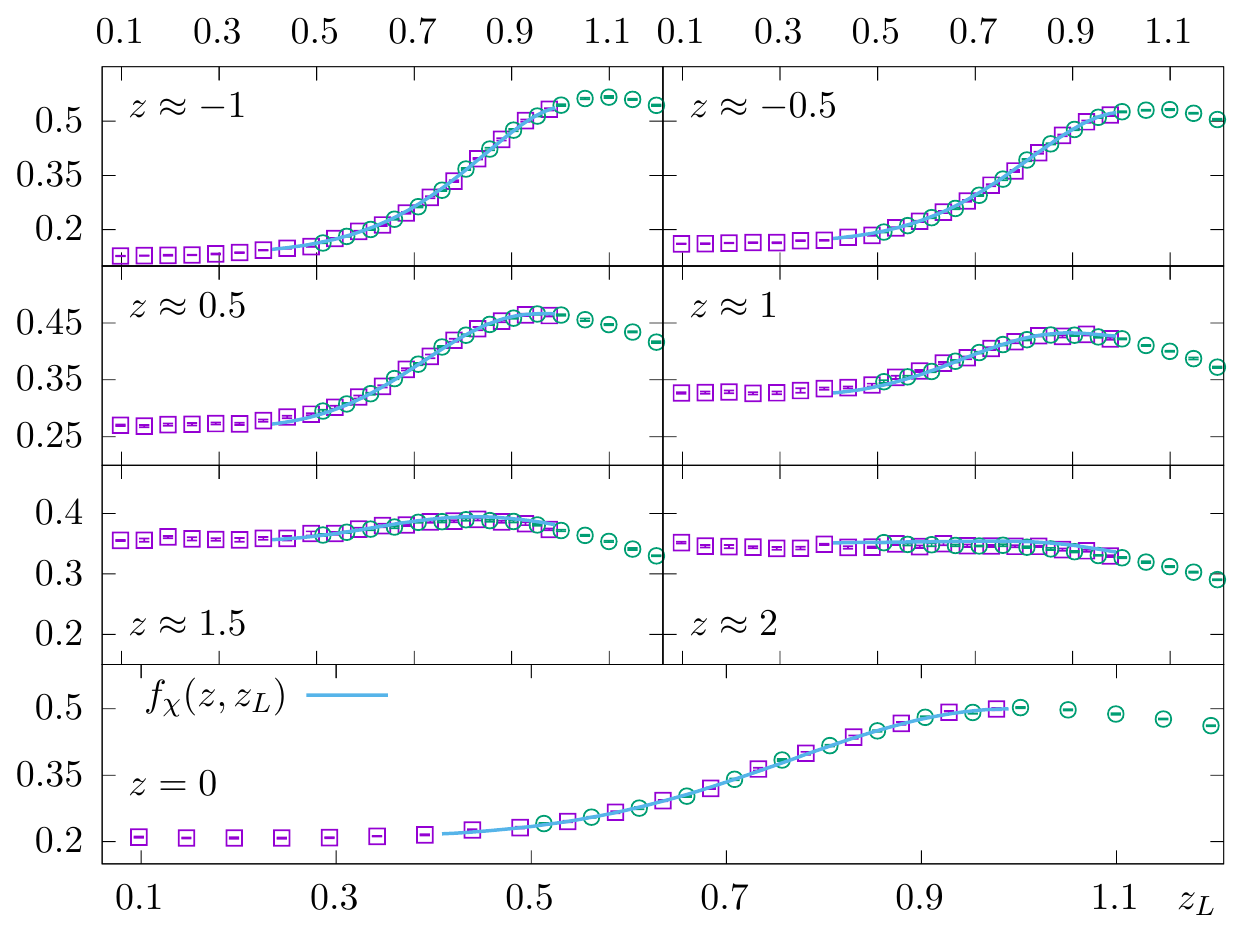}
\caption{Finite-size scaling functions $f_G(z,z_L)$ (top), $f'_G(z,z_L)$ (middle), and $f_\chi(z,z_L)$ (bottom) in the
$3$-$d$, $Z(2)$ (left) and $O(2)$
(right) universality classes. 
Shown are results for several fixed 
values $z\in [-1.0:2.0]$, obtained from calculations on lattices of size $L^3$, with $L= 48$ \,(green circles) and $96$ (purple squares),
in the range $z_L \in [0.1,1.2]$.
The light blue lines show results
of joined fits to all three 
scaling functions performed in the interval $z_L \in [0.4:1]$ (see Sec.~\ref{app:results}). As 
the values for $z$ and $z_L$ 
have been fixed a posterior, before
non-universal scale parameters have been
determined in the fits, we have labeled the different panels 
with approximate $z$-values. Their
actual values are (from lowest to highest) $Z(2)$: $z=-0.979, -0.490, 0, 0.490, 0.979, 1.469, 1.958$ and $O(2)$: $z=-1.051, -0.526, 0, 0.526, 1.051, 1.577, 2.010$.
}
\label{fig:fixedz}
\end{figure*}

\begin{figure*}[htb]
        \includegraphics[width=0.326\linewidth]{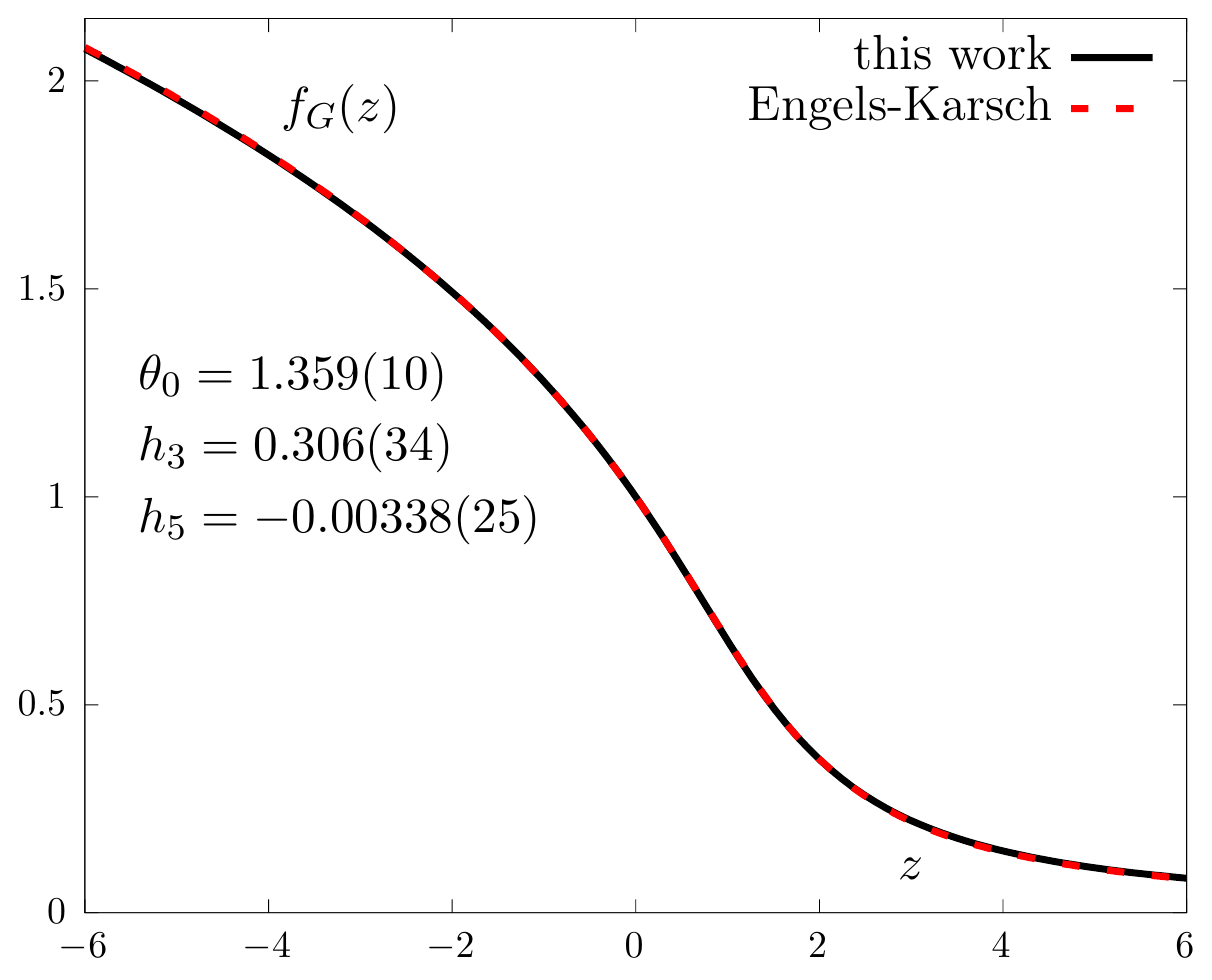}
        \includegraphics[width=0.326\linewidth]{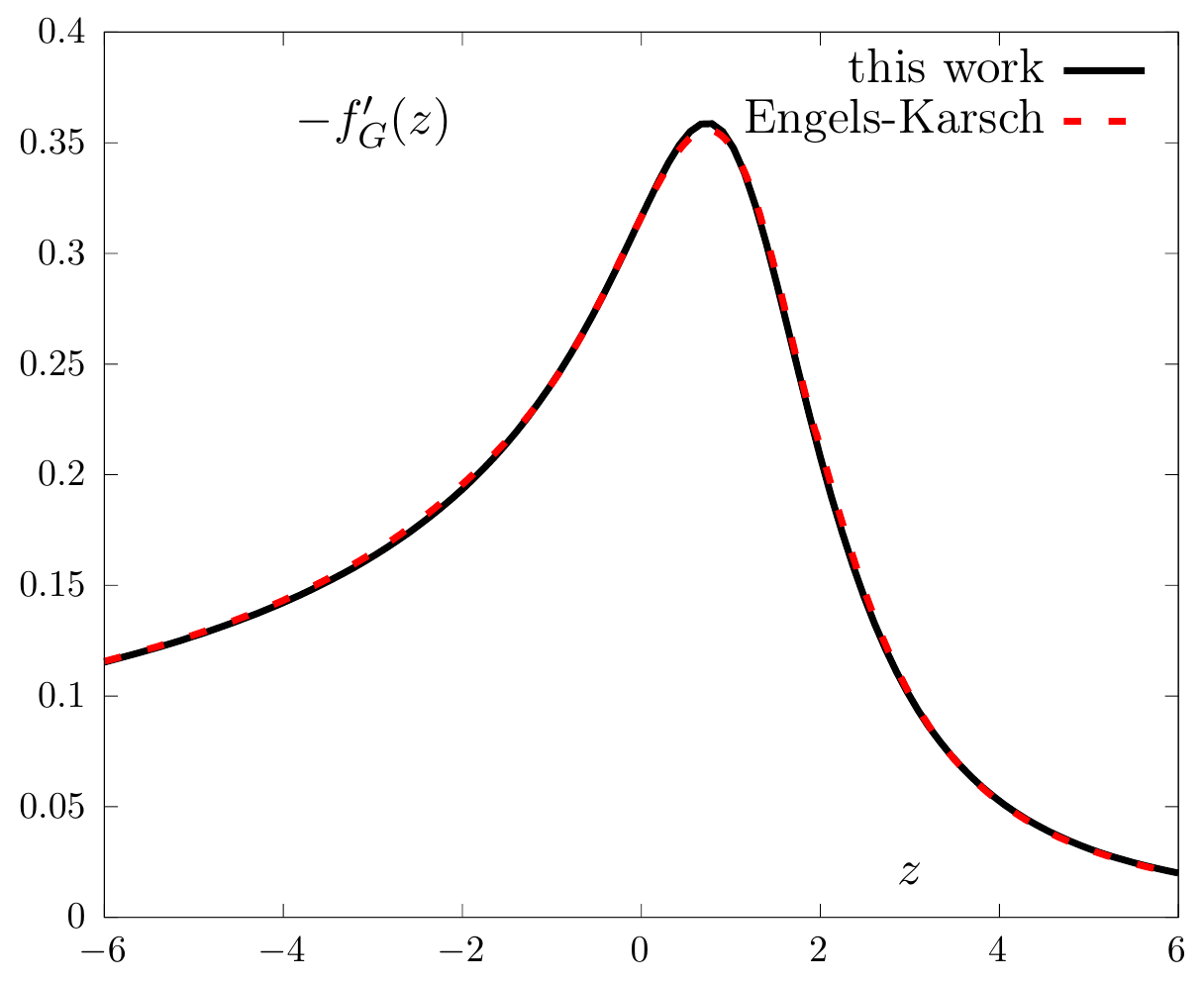}
        \includegraphics[width=0.326\linewidth]{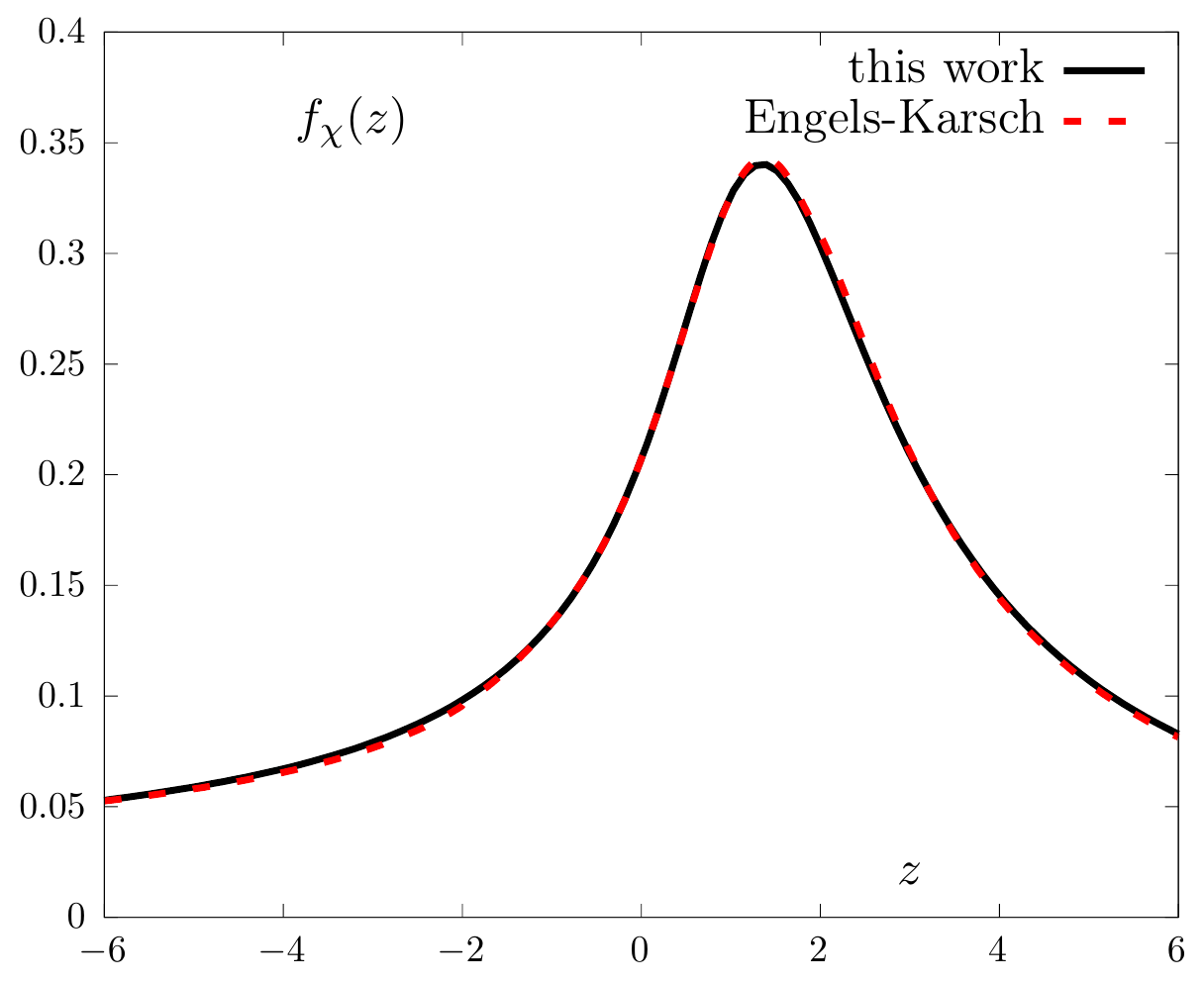}
    \caption{The $O(4)$ infinite volume scaling functions.
    The WG parameters have been obtained from a fit to $f_G$, while $f'_G$ and $f_\chi$ have been obtained from there. Dashed red lines show results from previous calculations \cite{Engels:2011km}.
    } \label{fig:O4-inf-vol}
\end{figure*}

In the following two sections, we will discuss results
for the infinite and finite volume scaling functions, respectively.
In order to judge which lattice sizes
and external field parameters are needed to get close to the infinite 
volume, universal scaling regime,
we first analyzed the $z_L$-dependence of the scaling functions 
at some fixed values of $z$ on different size lattices. 
The three scaling functions $f_G(z,z_L)$, $f'_G(z,z_L)$,
and $f_\chi(z,z_L)$ have been calculated at a few 
values of $z$ as functions of $z_L$. Using $z$ and $z_L$ as variables, 
of course, does require the determination of the non-universal scales 
$(t_0, H_0, L_0)$ which we are going to discuss in the next section 
and in Appendix~\ref{app:H0L0}.

Results from the calculations of the scaling 
functions for some fixed values of $z$,
performed on lattices of size $L^3$ with $L=48$ and $96$, are shown in
Fig.~\ref{fig:fixedz}. As can be seen, the finite-size
effects in all three scaling functions are almost negligible 
for $z_L < 0.4$. This is consistent
with findings obtained in calculations with the standard $O(4)$ model \cite{Engels:2014bra}
and can also be concluded from Fig.~\ref{fig:fG_compared}, shown in Appendix~\ref{app:H0L0}, where we 
compare results for $f_G(z=0,z_L)$ in different universality classes.

In the following, we thus use 
our numerical results for $z_L<0.4$
as approximation for infinite volume limit results.

\section{Infinite volume Scaling Functions}
\label{sec:scalingfct}
In our discussion of scaling functions in
the infinite volume limit, $z_L=0$, we suppress the second argument of the 
scaling functions, {\it i.e.} we introduce $f_{G}(z)\equiv f_{G}(z,0)$ 
and similarly for $f'_G(z)$ and $f_\chi(z)$.
These scaling functions have been determined previously using 
$\epsilon$-expansions \cite{Guida:1998bx} and 
perturbative field theoretic  approaches applied directly in 
$3$-dimensions \cite{Guida:1996ep,Zinn-Justin:1999opn,Campostrini:2000iw}, 
as well as in Monte Carlo (MC) simulations 
\cite{Engels:1999wf,Engels:2000xw,Engels:2002fi}. 
For the $Z(2)$ universality class it has been shown that the scaling functions, 
obtained in MC calculations, are in good agreement with
the Widom-Griffiths form \cite{Widom,PhysRev.158.176} using 
a resummed perturbative series for the order parameter obtained in
$3$-$d$ \cite{Zinn-Justin:1999opn}.
However, no parametrization based on MC results
has been given. The previous determination of the $O(2)$
scaling functions, using the WG ansatz \cite{Engels:2000xw}
has been performed using an
unimproved $O(2)$ Hamiltonian and, thus, had to take care of
corrections to scaling, which, in particular, made the determination 
of scaling functions in the symmetry broken regime difficult. 
Scaling functions for the $O(4)$ model, 
using Monte Carlo results obtained with the
standard, unimproved Hamiltonian (corresponding to $\lambda=\infty$), have
been presented in \cite{Engels:2011km}. In none of these cases has a 
parametrization of $O(N)$ scaling functions
been presented, which uses the WG form with only three free parameters.

\subsection{Widom-Griffiths form of scaling functions}
We present here a determination of the $Z(2)$ and $O(N)$ ($N=2,4$) 
scaling functions from Monte Carlo simulations. From our
new Monte Carlo results and those obtained in \cite{Engels:2011km}, 
we determine the parameters entering a parametrization of
scaling functions using the WG form of the order parameter scaling function \cite{Widom,PhysRev.158.176},
\begin{eqnarray}
M &=& m_0 R^{\beta}\theta\;, \label{W-G-M}\\
t &=& R(1-\theta^2)\; , \label{W-G-t}\\
h &=& h_0 R^{\beta\delta} h(\theta)\; , \label{W-G-h}
\end{eqnarray}
where ($R$,$\theta$) represents an alternate coordinate frame 
corresponding to the ($t,h$) plane \cite{PhysRevLett.22.606,PhysRevLett.23.1098}.
Aside from the normalization constants $m_0$ and $h_0$, 
this parametrization depends on a function $h(\theta)$, which
needs to be determined. For the case of $Z(2)$, it seems
that a Taylor series expansion up to ${\cal O}(\theta^5)$ is sufficient
\footnote{Note that $h(\theta)$ needs to be an odd function in $\theta$. Using the normalization constant $h_0$, one can assure that the coefficient of the leading order term is unity.} \cite{Zinn-Justin:1999opn},
while in the case of $O(N)$ one needs to take care explicitly of the
presence of Goldstone modes in 
the symmetry broken phase. This requires that $h(\theta)$
has a double zero at some $\theta_0>1$
\cite{Campostrini:2000iw,Campostrini:2002ky}.
We, thus, use the ansatz proposed
in \cite{Zinn-Justin:1999opn,Campostrini:2000iw},
\begin{eqnarray}
    \hspace*{-0.4cm}h(\theta)\hspace{-0.1cm} &=& \hspace{-0.1cm}
    (\theta +
    h_3 \theta^3 + h_5 \theta^5)
    \begin{cases}
    1\; ,
    &\hspace{-0.3cm}{\rm for}\;  Z(2)\; . \cr
    (1-\theta^2/\theta_0^2)^2 \; ,
    &\hspace{-0.3cm}{\rm for}\;  O(N)\; .
    \end{cases}
\label{hthetaZ2O2}
\end{eqnarray}

The normalization constants $m_0$ and $h_0$ are determined from the conditions in Eq.~\ref{Mconditions} which gives us
\begin{equation}
    m_0 = \frac{(\theta_0^2 - 1)^\beta}{\theta_0}\;, 
    \; h_0 = \frac{m_0^\delta}{h(1)}\,.
\end{equation}
where $\theta_0$ is the first positive zero of $h(\theta)$ in the $Z(2)$ universality class and the double zero of $h(\theta)$ in the $O(N)$ case.
Using Eqs.~\ref{zTzL} and \ref{fGzzL} and the above
relations for the normalization constants $m_0$ and $h_0$ one 
can establish the relation between
the WG form of the scaling function $f_G$ and the relation between the scaling variables,
$z=t/h^{1/\beta\delta}$ and $\theta$,
\begin{eqnarray}
f_G(z)\equiv  f_G(\theta (z)) &=&
 \theta \left( \frac{h (\theta)}{h (1)}\right)^{-1/\delta} \; ,
\label{W-G-fg} \\
z(\theta) &=&  \frac{1-\theta^2}{ \theta_0^2-1 }\theta_0^{1/\beta}
	\left( \frac{h(\theta)}{h(1)}\right)^{-1/\beta\delta}\; . 
	\label{W-G-z}
\end{eqnarray}
Obviously, $\theta=1$ corresponds 
to $z=0$ and 
$\theta=\theta_0$ corresponds to $z=-\infty$ and these, respectively, 
correspond to the normalization conditions for the scaling 
function $f_G$ in Eq.~\ref{fGconditions}.
Finally, $\theta=0$ corresponds to $z=\infty$. 
Using  Eqs.~\ref{W-G-fg} and \ref{W-G-z} 
we also obtain $f'_G(z)$ as,
\begin{equation}
    f'_G(z)\equiv \frac{{\rm d}f_G(\theta(z))}{{\rm d} z}=
    \frac{{\rm d}f_G}{{\rm d} \theta} \Big/
    \frac{{\rm d} z}{{\rm d}\theta}
    \label{fGpr}\; .
\end{equation}

The presence of Goldstone modes in the symmetry broken ($z<0$) phase of $O(N)$
symmetric models, gives rise to a 
distinctively different behavior of the $Z(2)$
and $O(N)$ model scaling function $f_G$ in the $z\rightarrow -\infty$ limit,
\begin{eqnarray}
\frac{f_G(z)}{(-z)^{\beta}} &=&
\begin{cases}
1+  
d_1^- (-z)^{-\beta\delta} + d_2^- (-z)^{-2\beta\delta}  \cr
 ~~ + {\cal O}((-z)^{-3\beta\delta})
\hspace{1.3cm},~{\rm for}\;\;\; Z(2)\; .
\cr
1 + d_1^- (-z)^{-\beta\delta/2} + d_2^- (-z)^{-\beta\delta}  \cr
 ~~ + {\cal O}((-z)^{-3\beta\delta/2}) 
\hspace{1.1cm},~{\rm for}\;\;\; O(N)\; .
\end{cases}
\label{fGzb}
\end{eqnarray}
One can arrive at the above asymptotic form for $f_G$ using Eqs.~\ref{W-G-fg} and \ref{W-G-z} with the ansatz for $h(\theta)$ in $Z(2)$ and $O(N)$ universality classes given in Eq.~\ref{hthetaZ2O2}.
Explicit expressions for
$d_1^-$ and $d_2^-$ in terms of the 
WG parameters $h_3, h_5$ and $\theta_0$
are given in Appendix~\ref{app:Widom}.

\begin{figure*}[t]
    \centering
        \includegraphics[width=0.45\linewidth]{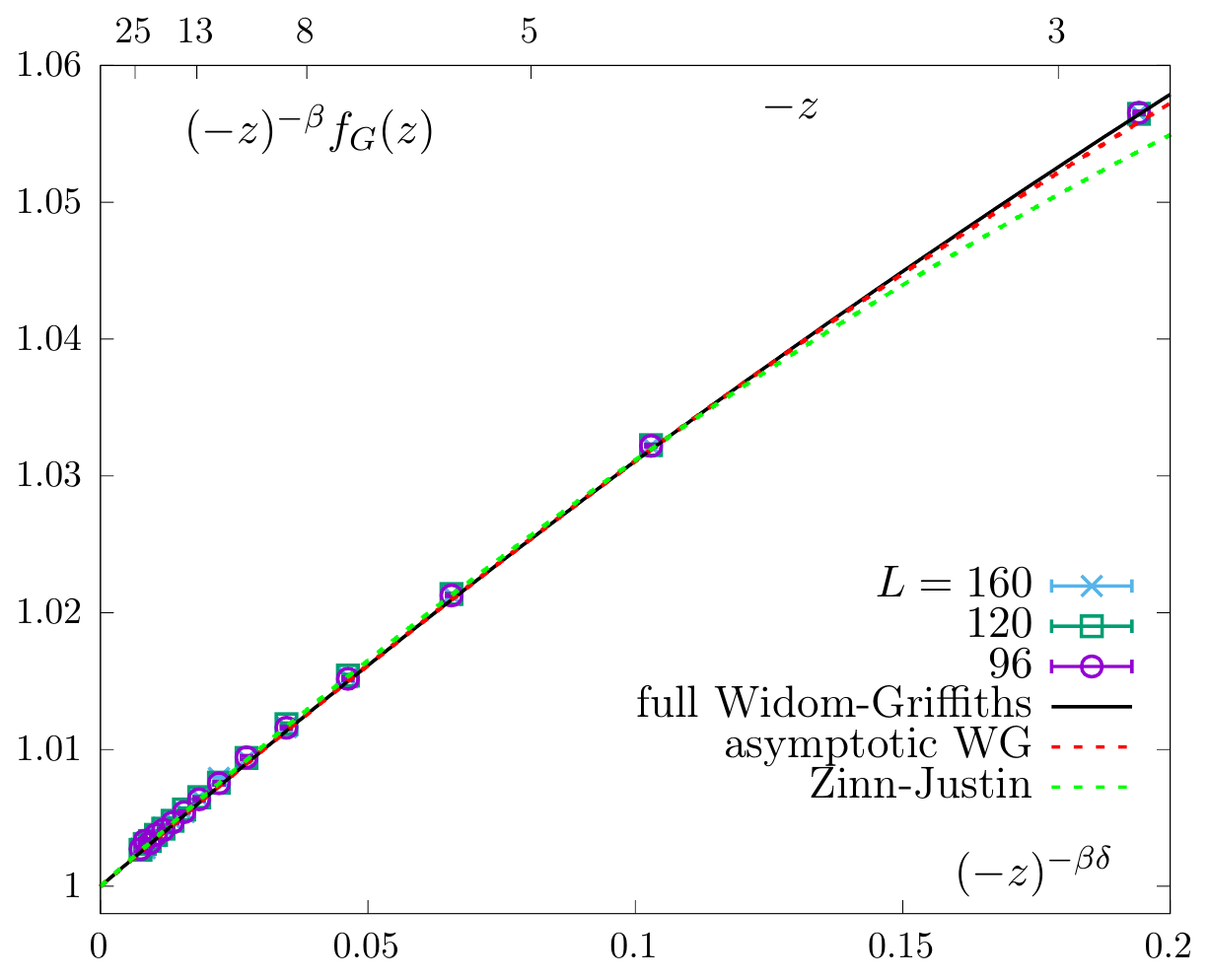}
    \includegraphics[width=0.45\linewidth]{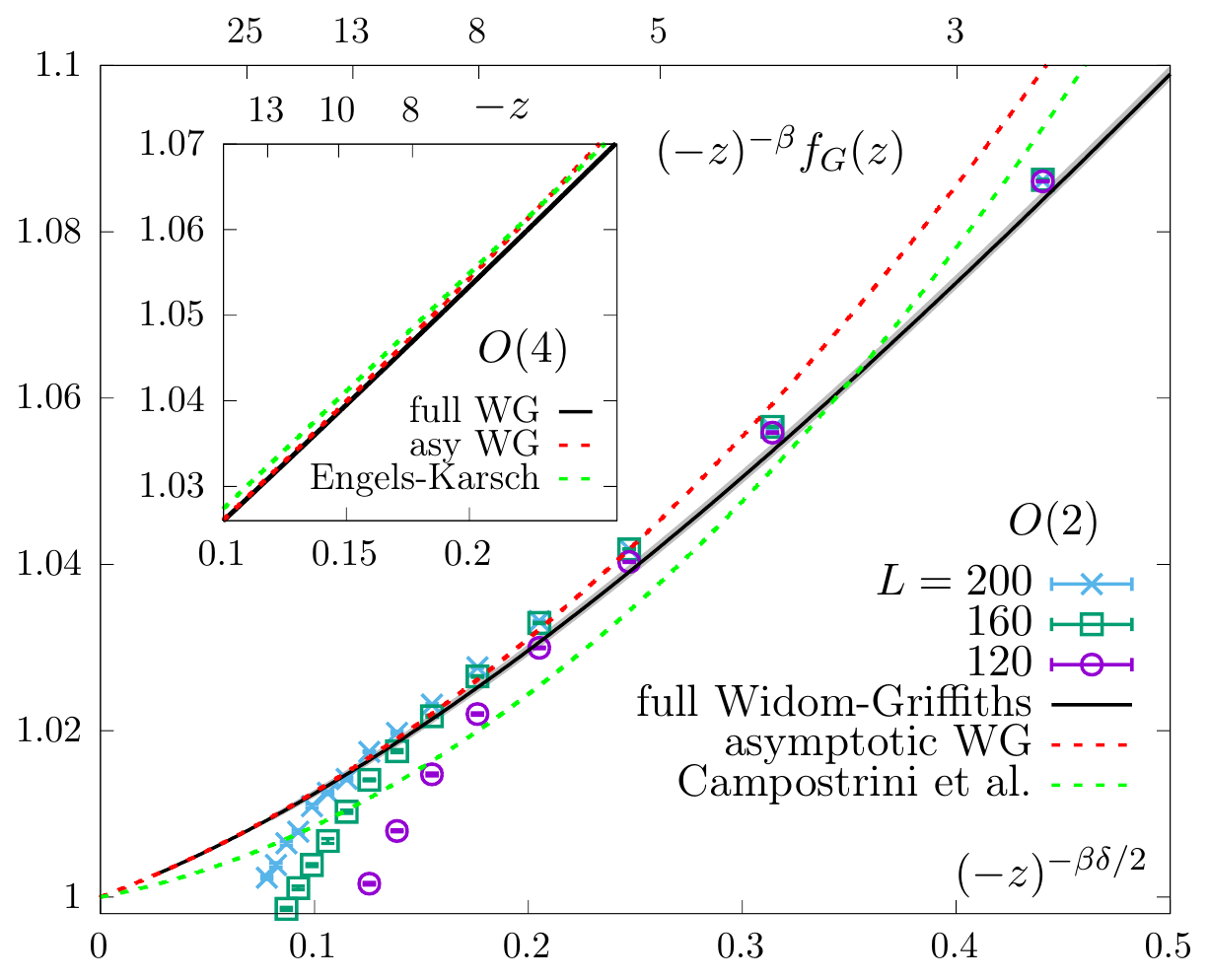}
    \caption{ The scaling function $f_G(z)$ in the $Z(2)$ (left) and 
    $O(2)$ (right) universality classes
    obtained from the order parameter $M$ using Eq.~\ref{fGzzL} 
    in the region of large, negative values of $z$. Monte Carlo 
    data have been obtained in simulations using the 
     $3$-$d$,
    $Z(2)$ model with $\lambda=1.1$ and 
the $O(2)$ model with $\lambda=2.1$. 
    All data are from simulations at $T/T_c=0.99$. 
    The large negative $z$ region of the $O(4)$ $f_G(z)$ 
    scaling function, obtained from our fit to results in \cite{Engels:2011km}, 
    is shown in the inset in the right figure. Solid lines 
    shown in the figures are based on fits using the Widom-Griffiths 
    form of the scaling functions and use also data outside the 
    parameter range shown here (see the text). For $O(2)$, we also 
    show an error band to the WG ansatz obtained from a bootstrap 
    analysis. The green dashed lines show the analytic results 
    obtained in the $Z(2)$ \cite{Zinn-Justin:1999opn} and $O(2)$
    \cite{Campostrini:2000iw} universality classes and MC fit results 
    obtained in the $O(4)$ \cite{Engels:2011km} universality class, 
    respectively. 
    The dashed red line shows the asymptotic expansion given in 
    Eq.~\ref{fGzb}.
}
    \label{fig:get_T0}
\end{figure*}

\subsection{Representation of \boldmath$O(4)$ scaling functions using the Widom-Griffiths form}
Although earlier parametrizations of the $O(4)$ scaling function $f_G(z)$, determined in Monte Carlo simulations \cite{Engels:1999wf}, made use of 
the Widom-Griffiths form, this was done only
to establish the behavior of $f_G(z)$ at 
large $|z|$. The region around $z=0$ has
been parametrized using polynomial ans\"atze.
A recent parametrization of the $O(4)$ scaling function used different fits in the 
small and large $z$ regions and obtained a 
parametrization that uses 14 parameters
\cite{Engels:2011km}. 

In order to establish the validity of the
WG form using an ansatz for the function $h(\theta)$ as suggested in \cite{Campostrini:2000iw} we re-parametrized 
the fit results presented in \cite{Engels:2011km}. 
We used the WG form for the $O(4)$ scaling functions as given in the
previous subsection and determined optimal parameters
$(\theta_0 , h_3 , h_5)$ in an interval around $z = 0$, i.e.,
we do not make use of the large $z$ behavior of the scaling function
given in Eq.~\ref{fGzb}. The structure of this asymptotic form is
implemented already in the Widom-Griffiths ansatz and
the expansion parameters $d_1^-$ and $d_2^-$ are determined
directly from $(\theta_0 , h_3 , h_5)$ (see Appendix \ref{app:Widom}), 
which can be determined from any set of $z$-values. 
We determined these parameters
in several intervals $[-z_{max} : z_{max} ] $ with $1 \leq z_{max} \leq 6$.
The resulting parameters are given in 
Table \ref{tab:O4_inf_vol_fit_paras}, where the errors quoted
there reflect the spread of results for ($\theta_0$ , $h_3$ , $h_5$)
obtained when varying $z_{max}$ . In Fig.~\ref{fig:O4-inf-vol}, 
we compare the scaling
functions $f_G (z)$, $f_G^\prime (z)$, and $f_\chi (z)$, obtained with
the WG ansatz using parameters given in Table \ref{tab:O4_inf_vol_fit_paras} 
to that obtained in \cite{Engels:2011km}.
As can be seen, we find excellent agreement.

Even though the scaling functions themselves are in good agreement and as such give 
consistent results for the positions $z_t$ and $z_p$ of the maxima of 
$-f'_G(z)$ and $f_\chi(z)$, we find different asymptotic behavior at large, 
negative $z$ as shown for the case of $f_G(z)$ in the inset in Fig.~\ref{fig:get_T0}(right). In \cite{Engels:2011km}, the sub-leading
asymptotic correction, $d_2^-$ has been found 
to vanish within errors, while we find 
$d_2^-\sim 0.1$. This difference, however, may 
not be too surprising, as the 
earlier results for the asymptotic expansion 
parameters $d_1^-$ and $d_2^-$ have been
obtained from fits in the interval $z\in [-10,-1]$. We will show in the next subsection that in the $O(2)$ case the asymptotic form is not yet valid in this $z$-range.

Given the good agreement between the 
WG parametrization of the $O(4)$ scaling functions and the earlier results based 
on a 14-parameter fit to MC data we find
it encouraging to analyze also the new 
Monte Carlo simulation results, obtained 
for the $3$-$d$, $Z(2)$ and $O(2)$ models
using a parametrization based on the WG ansatz.
\begin{table}[t]
\begin{minipage}[t]{0.45\textwidth}
    \centering
    \begin{tabular}{|c|r|r|}
    \hline
        \multicolumn{3}{|c|}{$O(4)$} \\
    \hline
    &Monte Carlo~~ & Monte Carlo~~~~ \\
    &WG-fit to \cite{Engels:2011km}~~ &  Engels-Karsch \cite{Engels:2011km}~\\
    \hline
    $h_3$ & 0.306(34)~ & --~~~~~~~~~~ \\
    $h_5$ & -0.00338(25)~ & --~~~~~~~~~~ \\
    $\theta_0$ & 1.359(10)~ & --~~~~~~~~~~ \\ \hline
    $d_1^-$ & 0.2481(20)~ & 0.2737(29)~ \\
    $d_2^-$ & 0.1083(50)~ & 0.0036(49)~ \\
    $z_t$ & 0.732(10)~ & 0.74(4)~ \\
    $z_p$ & 1.347(9)~ & 1.374(30)~ \\ \hline
    \end{tabular}
\end{minipage}
\vspace{0.2cm}

\caption{
    Fit parameters $h_3$, $h_5$, and $\theta_0$ for $O(4)$ infinite volume scaling functions in the Widom-Griffiths form appearing in the function $h(\theta)$
    introduced in Eq.~\ref{hthetaZ2O2}.
    In the lower part of the table, we give results for several universal constants computable from the WG parametrization.
}
\label{tab:O4_inf_vol_fit_paras}
\end{table}

\subsection{Representation of \boldmath$Z(2)$ and $O(2)$ scaling functions using the Widom-Griffiths form}

In order to use Eqs.~\ref{W-G-fg} and \ref{W-G-z} in determination of the  
scaling functions $f_G, f'_G$, and $f_\chi$ from MC results, 
as given in Eqs.~\ref{fGzzL}, \ref{fGprzzL} and \ref{fchizzL},
one still needs to determine the non-universal scale parameters $(t_0, H_0, L_0)$.
The non-universal scales $H_0$ and $L_0$ can be determined from the finite-size dependence of $f_G(z,z_L)$ at $T_c$, 
{\it i.e.}, at $z=0$. 
We present a determination of these two scales in Appendix~\ref{app:H0L0}. 
Once they have been determined from our 
results on different size lattices,
the scale parameter $t_0$ can be determined
from the asymptotic behavior of $f_G(z)$ in
the limit $z\rightarrow -\infty$.
Using Eq.~\ref{Msig}, the second normalization condition in Eq.~\ref{Mconditions} and
writing $z=z_0 z_b$ with $z_0=H_0^{1/\beta\delta}/t_0$, we
obtain $t_0$ from
\begin{equation}
    t_0^{-\beta}= \lim_{z_b\rightarrow -\infty} (-z_b)^{-\beta} H^{-1/\delta} M(T,H,\infty) \; .
    \label{Minfty}
\end{equation}

As this equation relates the scale $t_0$ to
observables calculated in the infinite volume
limit, its determination  can directly be
incorporated into fits which we perform in the infinite volume limit for the 
determination of the scaling functions.
We obtain $t_0$ and the parameters $(h_3,h_5,\theta_0)$ defining $h(\theta)$
using simultaneous fits to the scaling functions $f_G(z)$, $f'_G(z)$, and
$f_\chi(z)$ defined in Eqs.~\ref{fGzzL}-\ref{fchizzL}.
While in the parametrization of $Z(2)$ 
scaling functions $\theta_0$
is a function of $(h_3,h_5)$, it is 
an additional free parameter in the $O(N)$ case. 

\begin{table}[t]
\begin{minipage}[t]{0.45\textwidth}
    \centering
    \begin{tabular}{|c|r|r|}
    \hline
        \multicolumn{3}{|c|}{$Z(2)$} \\
    \hline
    &Monte Carlo~ & $3$-$d$ perturbative \\
    &(this work)~~ & expansion  \cite{Zinn-Justin:1999opn}~~~\\
    \hline
$h_3$ & -0.6274(26)~ & -0.76201(36)~ \\
$h_5$ & 0.05360(12)~ & 0.00804(11)~ \\ \hline
$\theta_0$ & 1.3797(24)~ & 1.15369(17)~ \\
$d_1^-$ & 0.33553(83)~ & 0.348329(13)~ \\
$d_2^-$ & -0.2466(71)~ & -0.368672(53)~ \\
$z_t$ & 0.8961(10)~ & 0.8578(3)~ \\
$z_p$ & 1.9770(23)~ & 1.9863(3)~ \\ \hline
    \end{tabular}
\end{minipage}
\vspace{0.2cm}

\begin{minipage}[t]{0.45\textwidth}
    \centering
    \begin{tabular}{|c|r|r|}
    \hline
    \multicolumn{3}{|c|}{$O(2)$} \\
    \hline
    &~Monte Carlo~~ & $3$-$d$ improved High-T~ \\
    &(this work)~~ & expansion \cite{Campostrini:2000iw}~~~~~\\
    \hline
$h_3$ & 0.162(20)~ & 0.0758028~ \\
$h_5$ & -0.0226(18)~ & 0~ \\
$\theta_0$ & 1.610(14)~ & 1.71447~ \\ \hline
$d_1^-$ & 0.0969(38)~ & 0.04870~ \\
$d_2^-$ & 0.2925(61)~ & 0.36632~ \\
$z_t$ & 0.7991(96)~ & 0.8438~  \\
$z_p$ & 1.6675(68)~ & 1.7685~  \\ \hline
    \end{tabular}
\end{minipage}
\caption{{\it Top:}
Fit parameters $h_3$ and $h_5$ for $Z(2)$ infinite volume scaling functions in the Widom-Griffiths form appearing in the function $h(\theta)$
    introduced in Eq.~\ref{hthetaZ2O2}.
{\it Bottom:}
    Fit parameters $h_3$, $h_5$ and $\theta_0$ for $O(2)$ infinite volume scaling functions in the Widom-Griffiths form appearing in the function $h(\theta)$
    introduced in Eq.~\ref{hthetaZ2O2}.
    In the lower part of both
    tables, we give results for several universal constants computable from the WG parametrization.
}
\label{tab:Z2-O2_inf_vol_fit_paras}
\end{table}

Goldstone modes dominate
finite-size effects at large, negative values of $z$, which are quite different in $O(N)$ universality classes
from those in the $Z(2)$ case.
In the $O(2)$ universality class finite-size effects grow rapidly with decreasing values
of $z$. This is evident from Fig.~\ref{fig:get_T0}, where we 
show results for $(-z)^{-\beta}f_G(z)$ in the 
region $z<-2$. The figure shows that we had to 
perform MC calculations on rather large lattices to extract the scale 
parameter $t_0$ from the asymptotic
behavior of the order parameter in the symmetry broken phase. In our simulations of the $O(2)$ model, lattices of size $L^3$ with $L=200$ were 
needed to reach the region $z\le -10$ without suffering from finite-size effects.
In the case of $Z(2)$, lattices with 
$L=96$ were already sufficient to 
perform calculations in a region down to values $z\simeq -20$ without observing a significant finite-size dependence in 
our results. 

\begin{figure*}[htb]
        \includegraphics[width=0.426\linewidth]{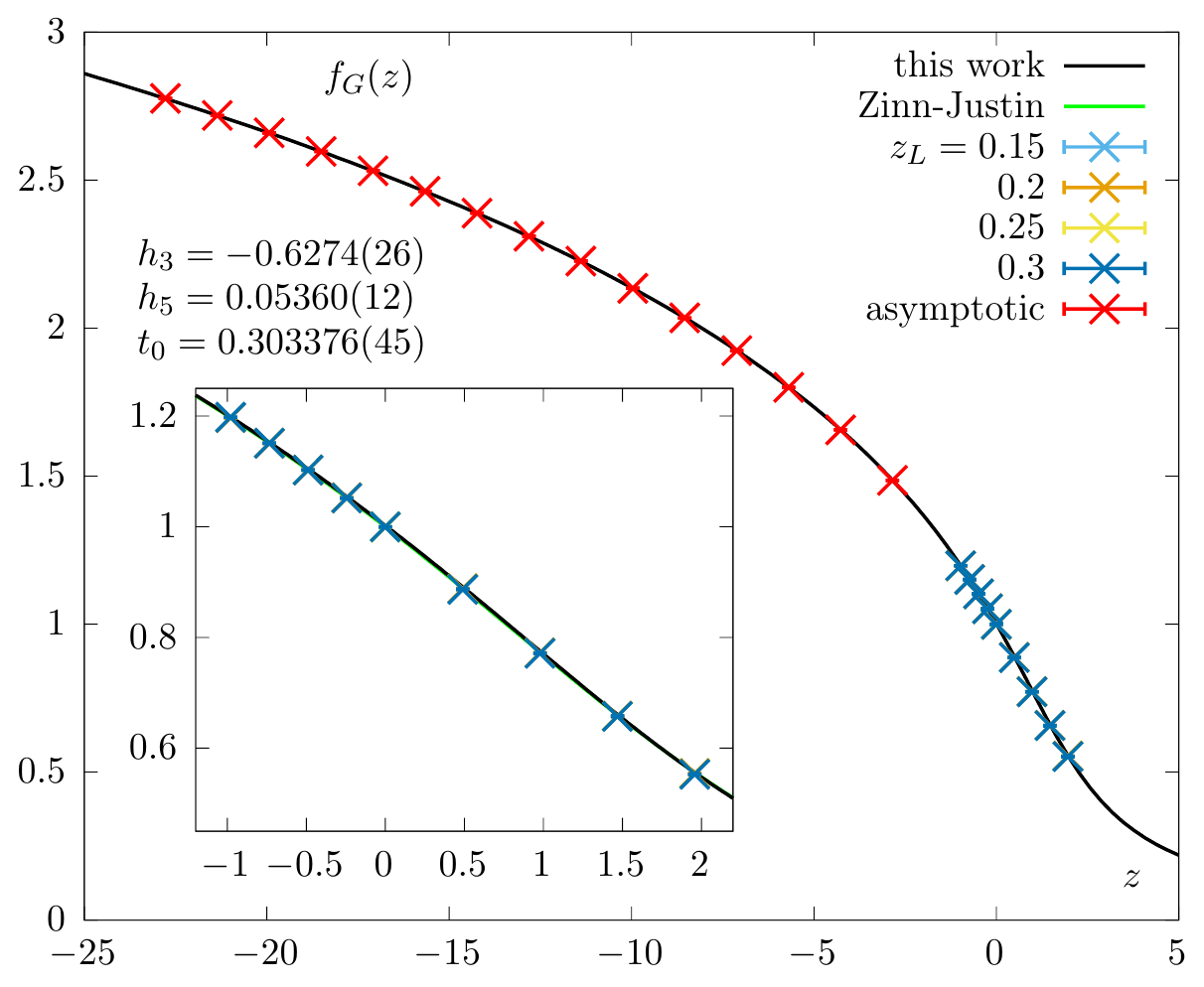}\hspace{0.7cm}
        \includegraphics[width=0.426\linewidth]{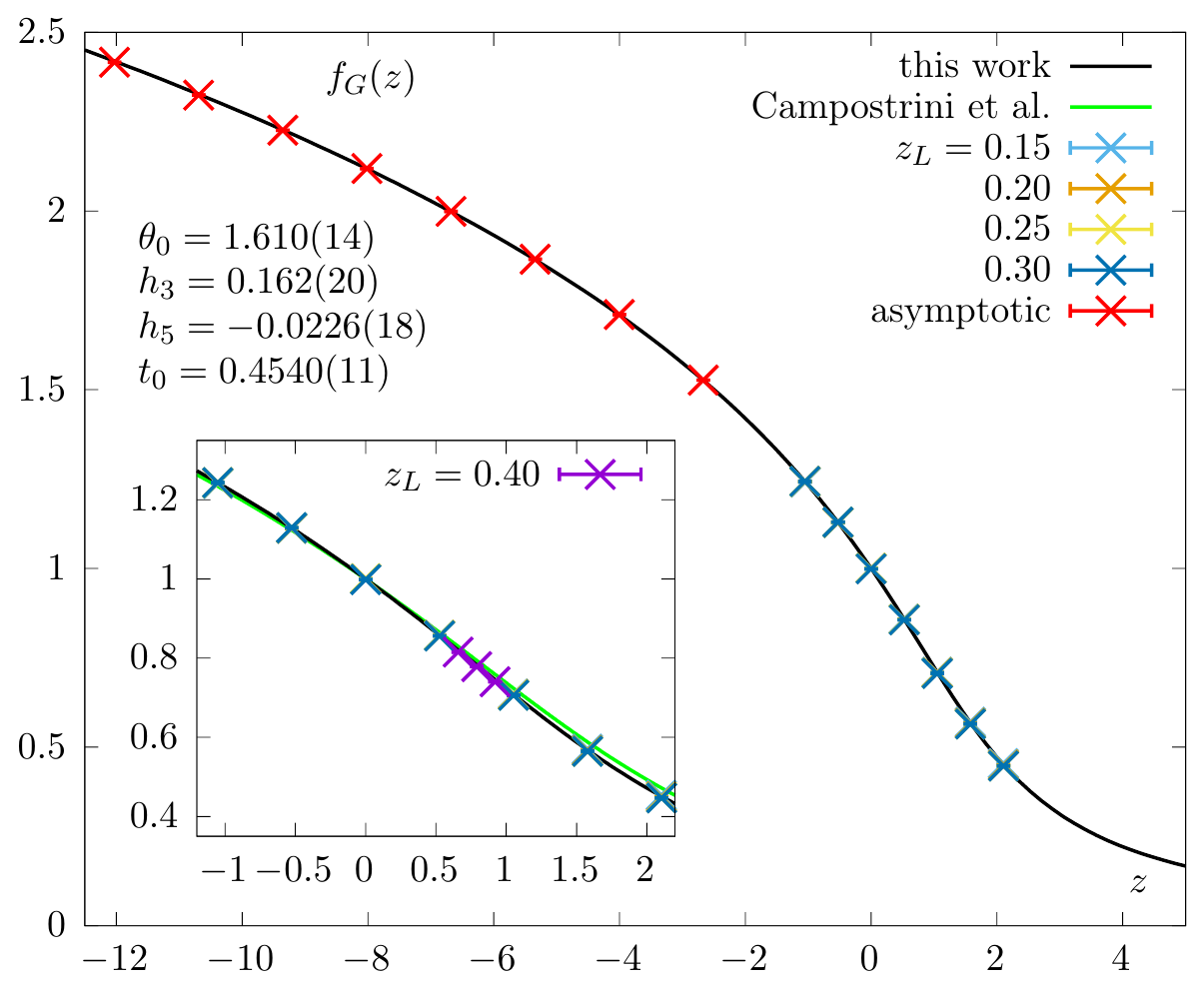}
        \includegraphics[width=0.426\linewidth]{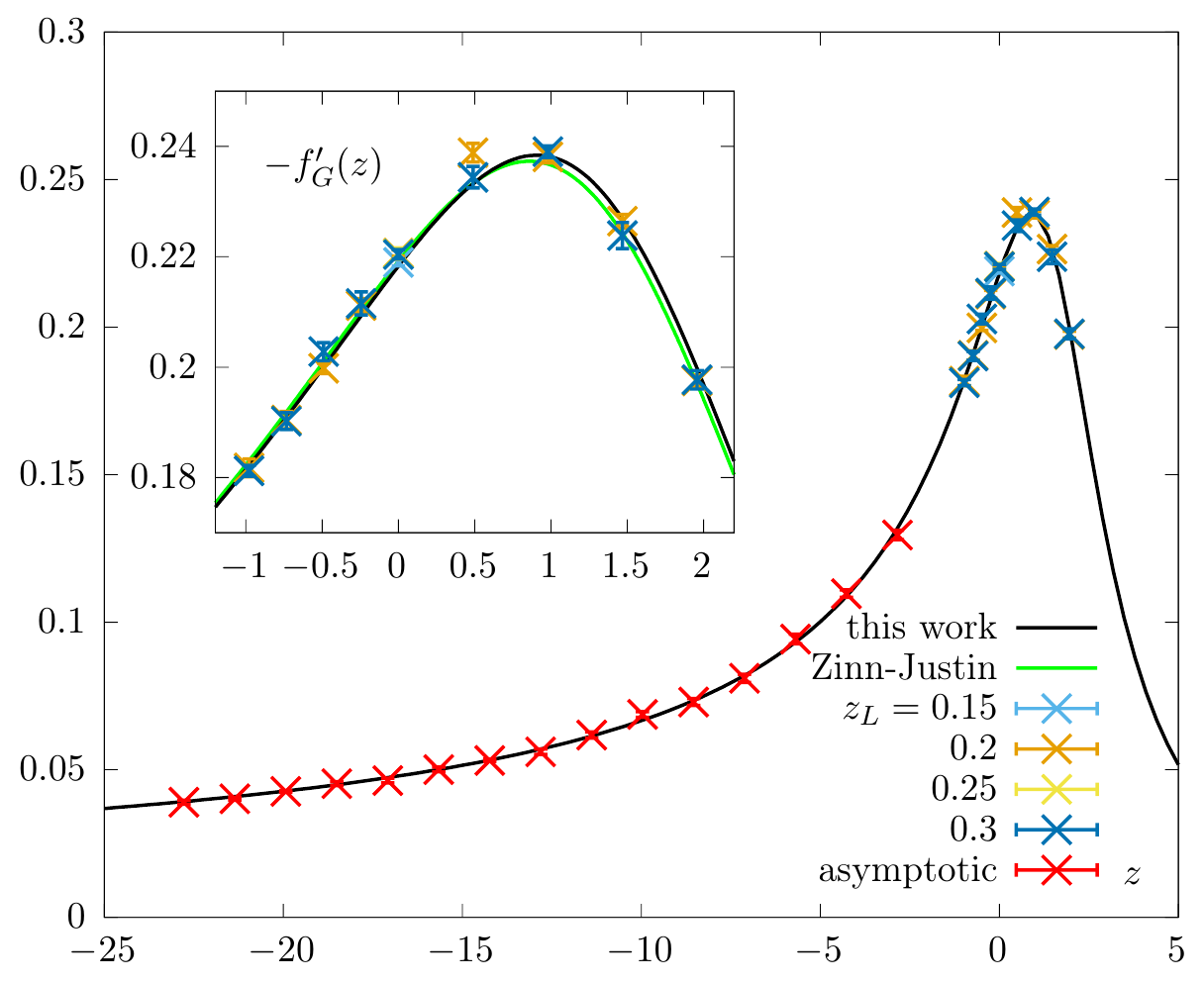}\hspace{0.7cm}
        \includegraphics[width=0.426\linewidth]{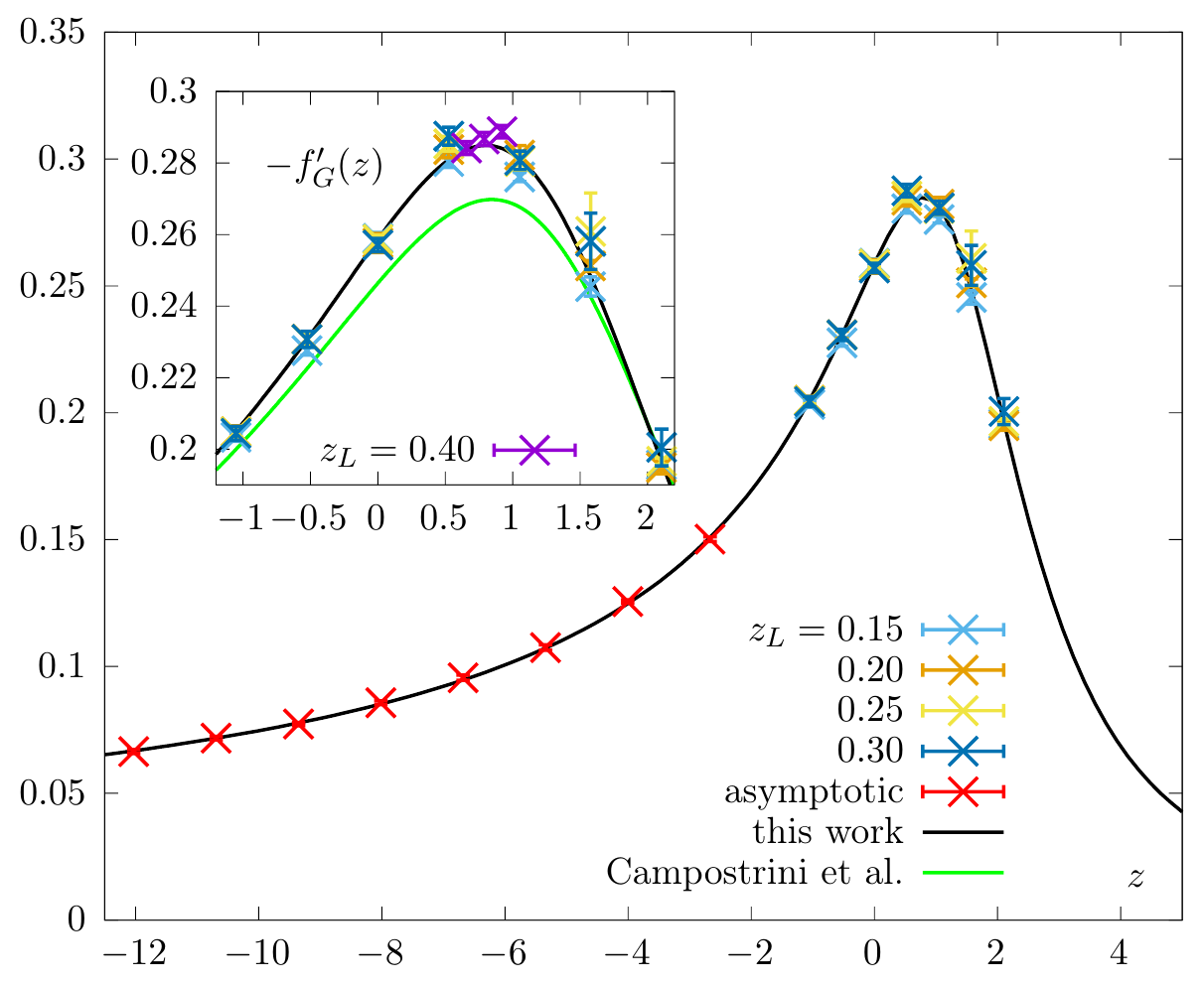}
        \includegraphics[width=0.426\linewidth]{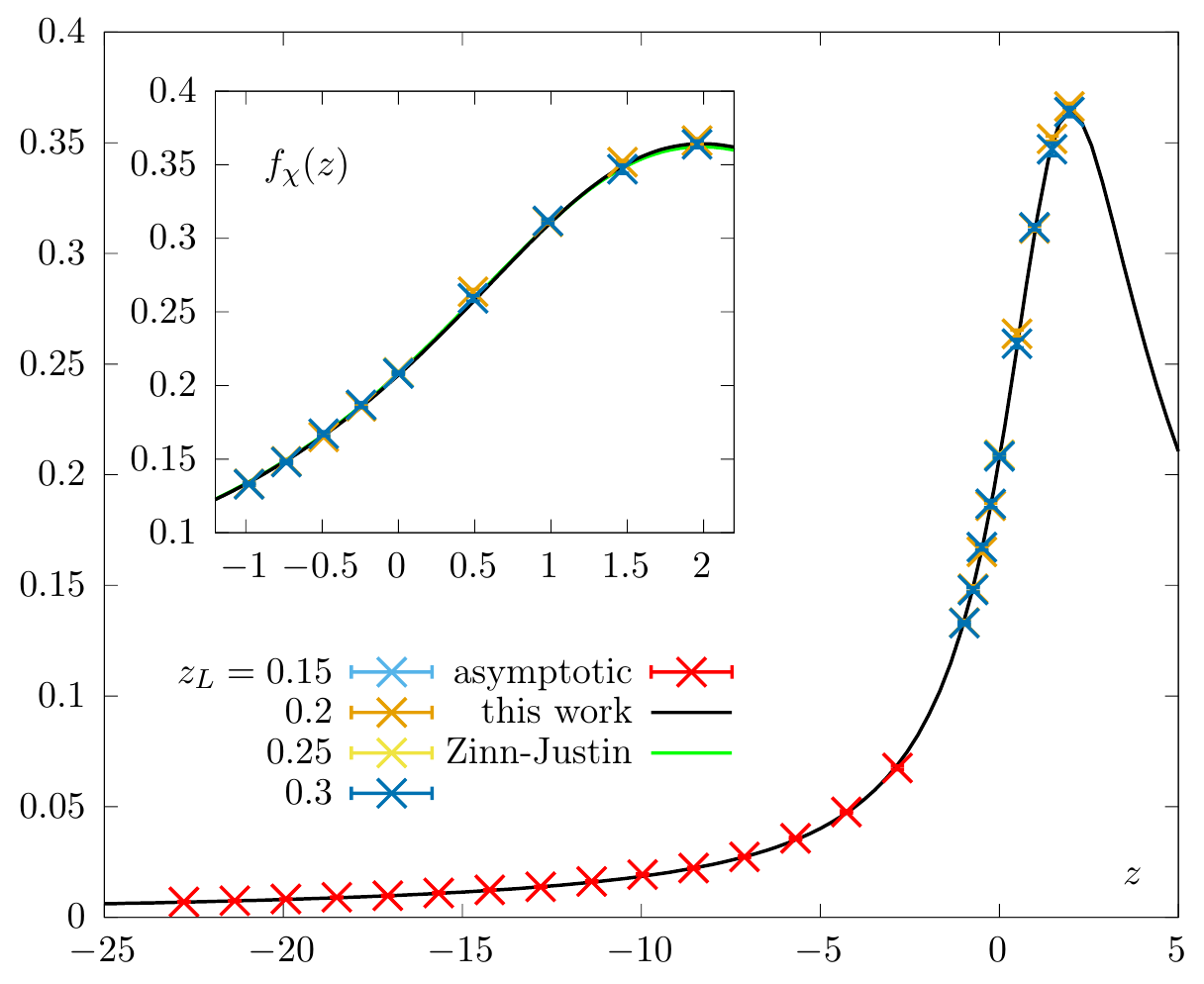}\hspace{0.7cm}
        \includegraphics[width=0.426\linewidth]{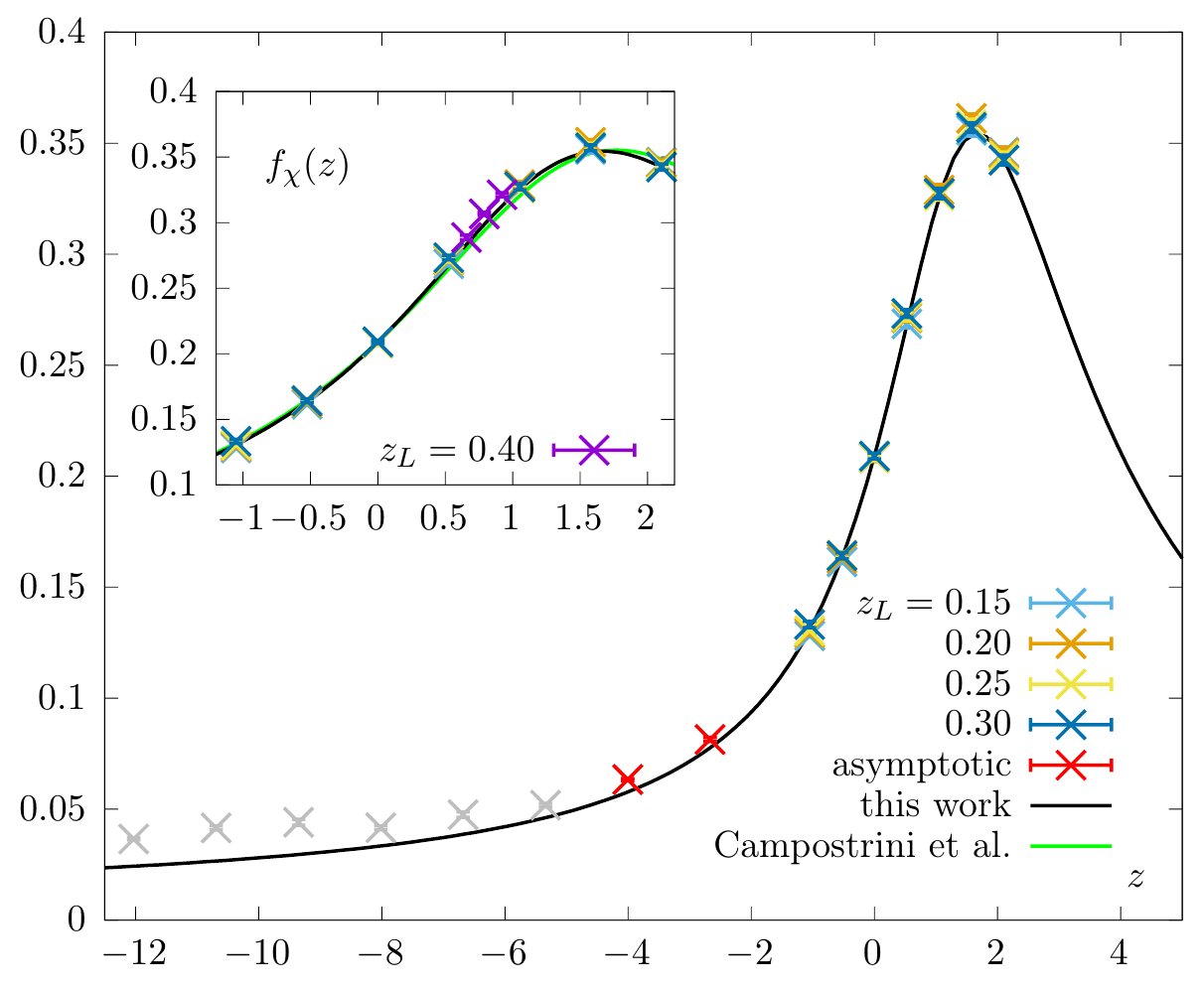}
    \caption{The $Z(2)$ (left column) and $O(2)$ (right column) infinite volume 
    scaling functions. Shown are Monte Carlo results obtained on lattices of size 
    $L^3$ with $L=96$ for $|z|\le 2.1$, $L=96,120,160$ ($Z(2)$, only $L=160$ is shown) 
    and $L=200$ ($O(2)$) for $z<-2.5$.
    All fits are joint fits to data for $f_G$, $f'_G$ and $f_\chi$, close to $z=0$ 
    and in the large, negative $z$ regime. The gray data were not included in the fit. 
    Green lines in the insets show results from analytic calculations
    \cite{Zinn-Justin:1999opn} ($Z(2)$, left) and \cite{Campostrini:2000iw} 
    ($O(2)$, right).
    } \label{fig:Z2-O2-inf-vol}
\end{figure*}

For $z>-2$, it was sufficient to perform calculations on lattices with
$L=48-120$. For $z<-2$, however, we also performed
calculations with $L=160$ and $200$ 
for the $O(2)$ model and $L=160$ in the case of $Z(2)$.
The statistics collected in all parameter
ranges are given in Tables~\ref{tab:statistics}-\ref{tab:statistics_T0}.

Our Monte Carlo results obtained for the 
$3$-$d$, $Z(2)$ and $O(2)$ models in the 
large volume limit, $z_L\le 0.35$,
are shown in Fig.~\ref{fig:Z2-O2-inf-vol}.
We performed joint fits using data in the region $z_L\le 0.35$ as approximation for 
the infinite volume limit.
All three scaling functions have then been obtained from joint fits to the WG form in the 
range $z\in [-23:2]$ for $Z(2)$ and 
$z\in [-12:2]$ for the $O(2)$ model\footnote{Note that the exact fit range is determined only a posterior, once
the scale parameter $t_0$ has been obtained in our fits.}.

We summarize results for the non-universal scale
parameters $(t_0,H_0,L_0)$, determined by us, in Table~\ref{tab:parameter}.
In Table~\ref{tab:Z2-O2_inf_vol_fit_paras}
we give all universal fit parameters
entering the definition of $h(\theta)$, and compare with results obtained in
$3$-$d$ analytic calculations \cite{Zinn-Justin:1999opn,Campostrini:2000iw}.
In the top section of the two tables, we 
give the parameters
$(h_3,h_5)$, entering fits performed for scaling functions in the 
$Z(2)$ universality class, and  $(\theta_0,h_3,h_5)$ in the $O(2)$ 
case. Results for the non-universal fit parameter $t_0$, obtained in the
same fits, are given in Table~\ref{tab:parameter}.
The bottom part of the tables gives
some universal constants derived from 
the Widom-Griffiths form of the scaling functions by using, on the one hand, 
the results from fits to our MC data and, on the other hand, the perturbative 
results for $h(\theta)$ and $\theta_0$ as input.

Aside from the parameters $d_1^-$ and
$d_2^-$ controlling the asymptotic 
behavior of $f_G(z)$ at large, negative $z$ (Eq.~\ref{fGzb}) we 
also give there the universal 
constants $z_p$ and $z_t$, which 
are the $z$-values at the 
maxima of $f_\chi(z,0)$ and $-f'_G(z,0)$, respectively. 

For the ratio of $z_p$ and $z_t$, determining pseudo-critical temperatures in the $Z(2)$ and 
$O(N)$ universality classes, we find
\begin{equation}
    \frac{z_p}{z_t}=
    \begin{cases}
    2.21(1) &, ~ Z(2)\; , \\
    2.09(2) &, ~O(2) \; , \\
    1.84(1) &, ~O(4) \; .
    \end{cases}
\end{equation}

In Fig.~\ref{fig:get_T0}, we compared the MC 
results  for $f_G(z)$ at large, negative 
values of $z$, {\it i.e.} for $z< -2$, with 
the WG form of the scaling function, given in Eq.~\ref{W-G-fg}, as well as with the
asymptotic form given in Eq.~\ref{fGzb}.
As can be seen, in the $Z(2)$ universality
class, the asymptotic expansion using the 
first two sub-leading corrections, gives
a good approximation to the  full WG form, in 
almost the entire region, $z<-2$. 
In the $O(2)$ and $O(4)$ universality classes, however, the first two sub-leading corrections 
agree with the full WG form only for $z<-(8-10)$ 
as can be seen in Fig.~\ref{fig:get_T0}(right) for the $O(2)$ case 
and in the inset in Fig.~\ref{fig:get_T0}(right) for the $O(4)$ case.

Also shown in Fig.~\ref{fig:get_T0} 
are the results obtained from the $3$-$d$, analytic 
calculations \cite{Zinn-Justin:1999opn, Campostrini:2002ky}. While in the $Z(2)$ case differences
are insignificant, they clearly are visible
in the $O(2)$ case. 
However, in the asymptotic regime both parametrizations of the WG form differ
by less than 1\%. 
We observed the largest differences in the vicinity
of the maximum of $-f'_G(z)$, where deviations
between the analytic and MC calculation amount
to about 5\%. This is apparent from the insets
shown in Fig.~\ref{fig:Z2-O2-inf-vol}.

\begin{figure*}[htb]
        \includegraphics[width=0.43\linewidth]{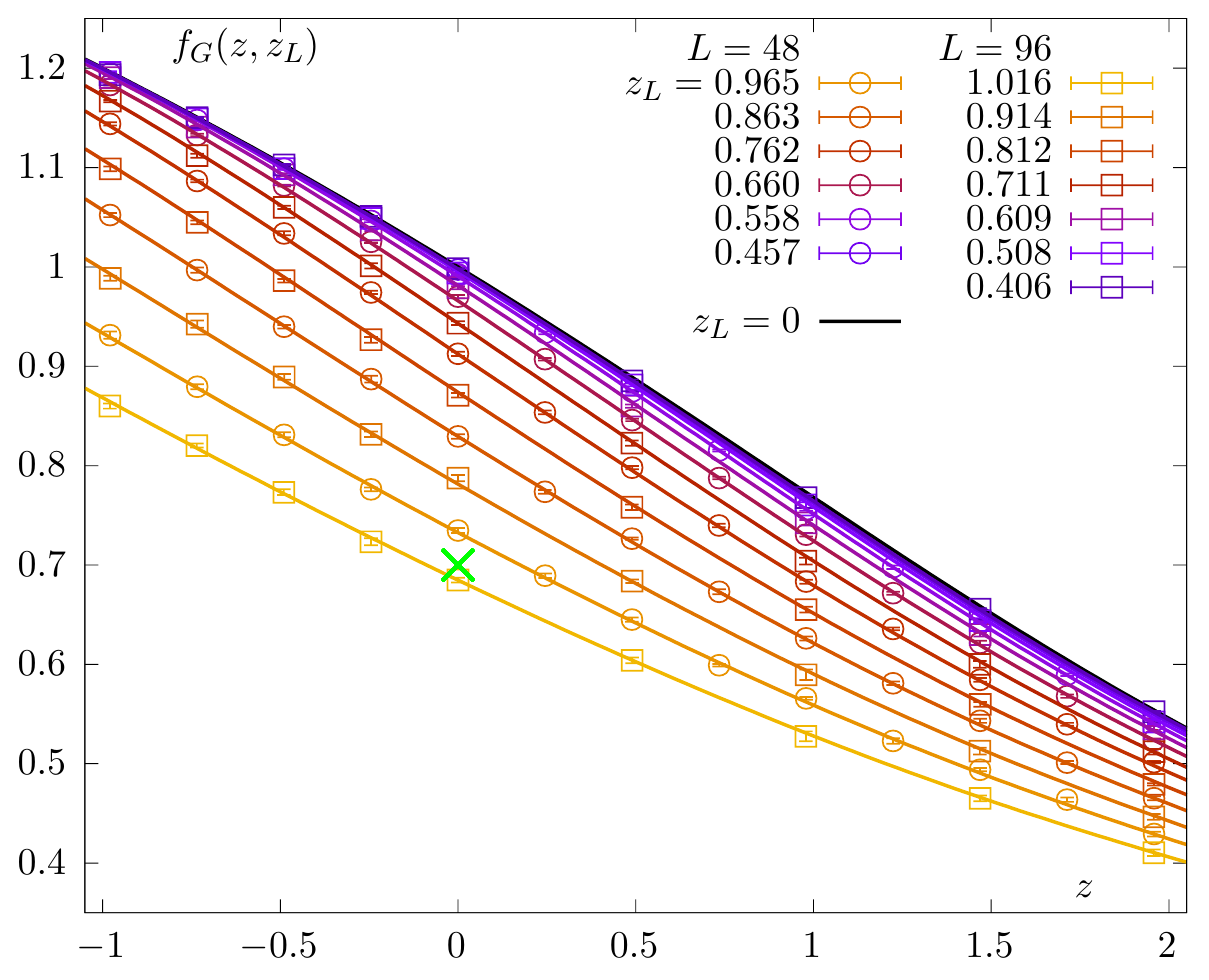}\hspace{0.7cm}
        \includegraphics[width=0.43\linewidth]{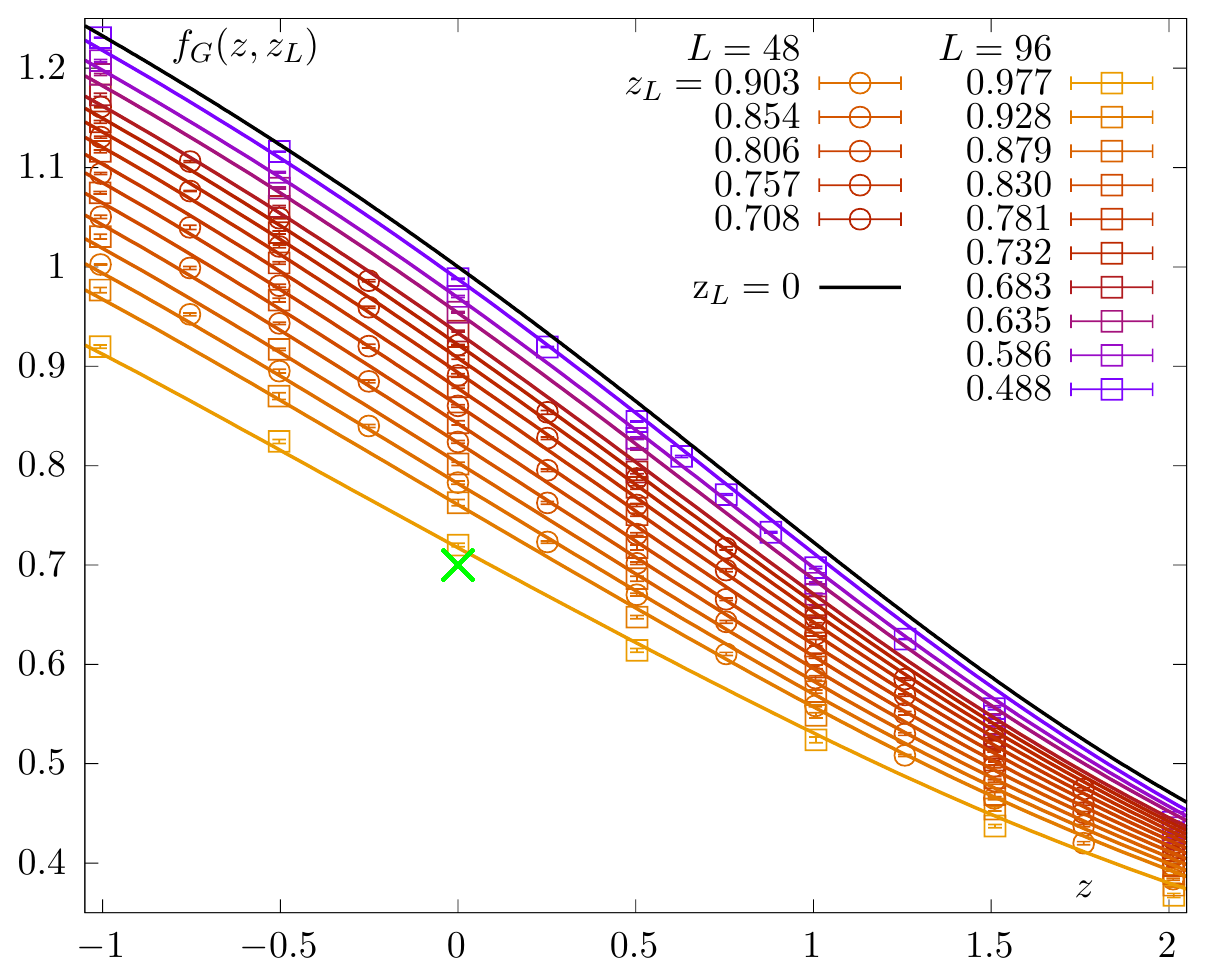}
        \includegraphics[width=0.43\linewidth]{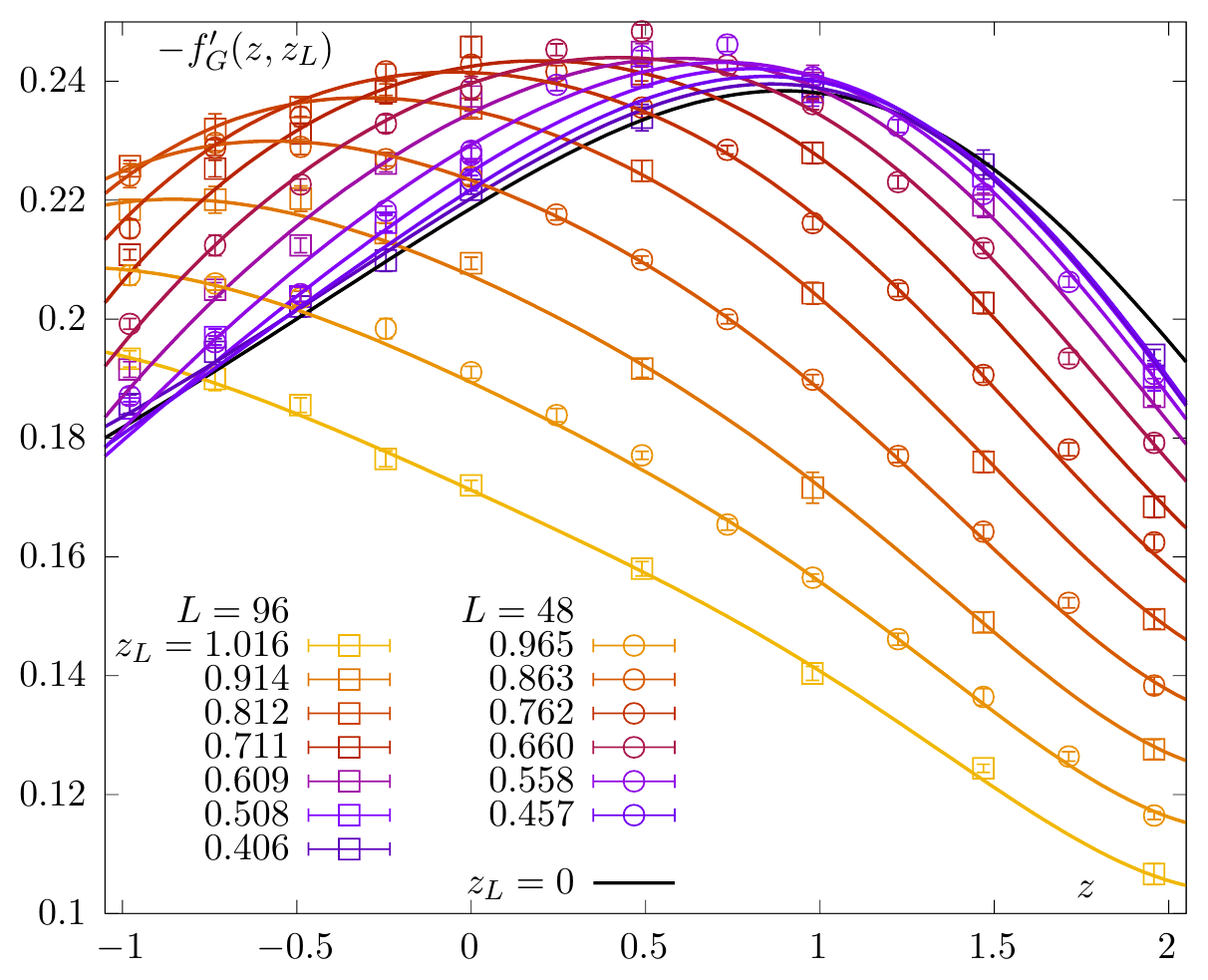}\hspace{0.7cm}
        \includegraphics[width=0.43\linewidth]{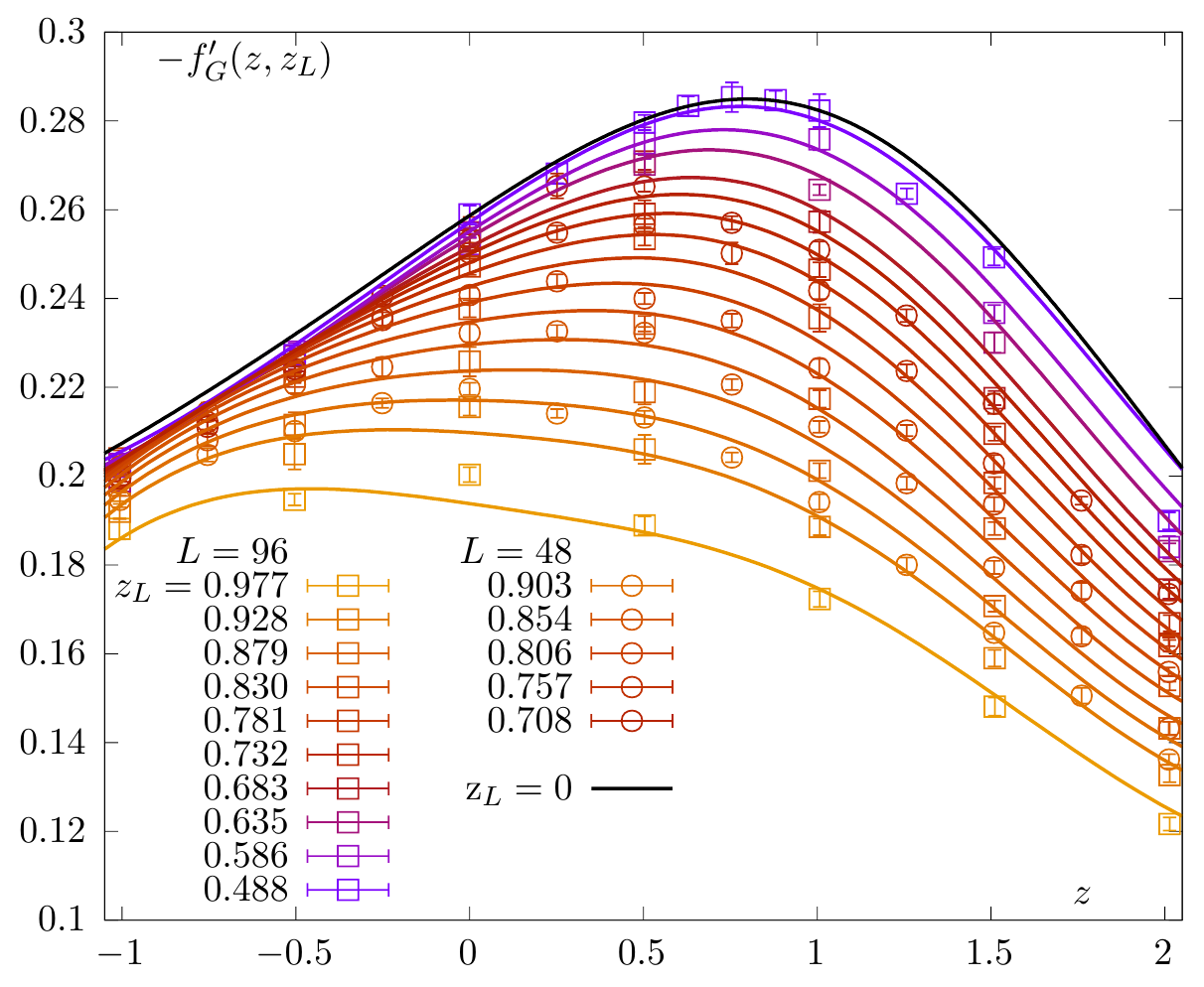}
        \includegraphics[width=0.43\linewidth]{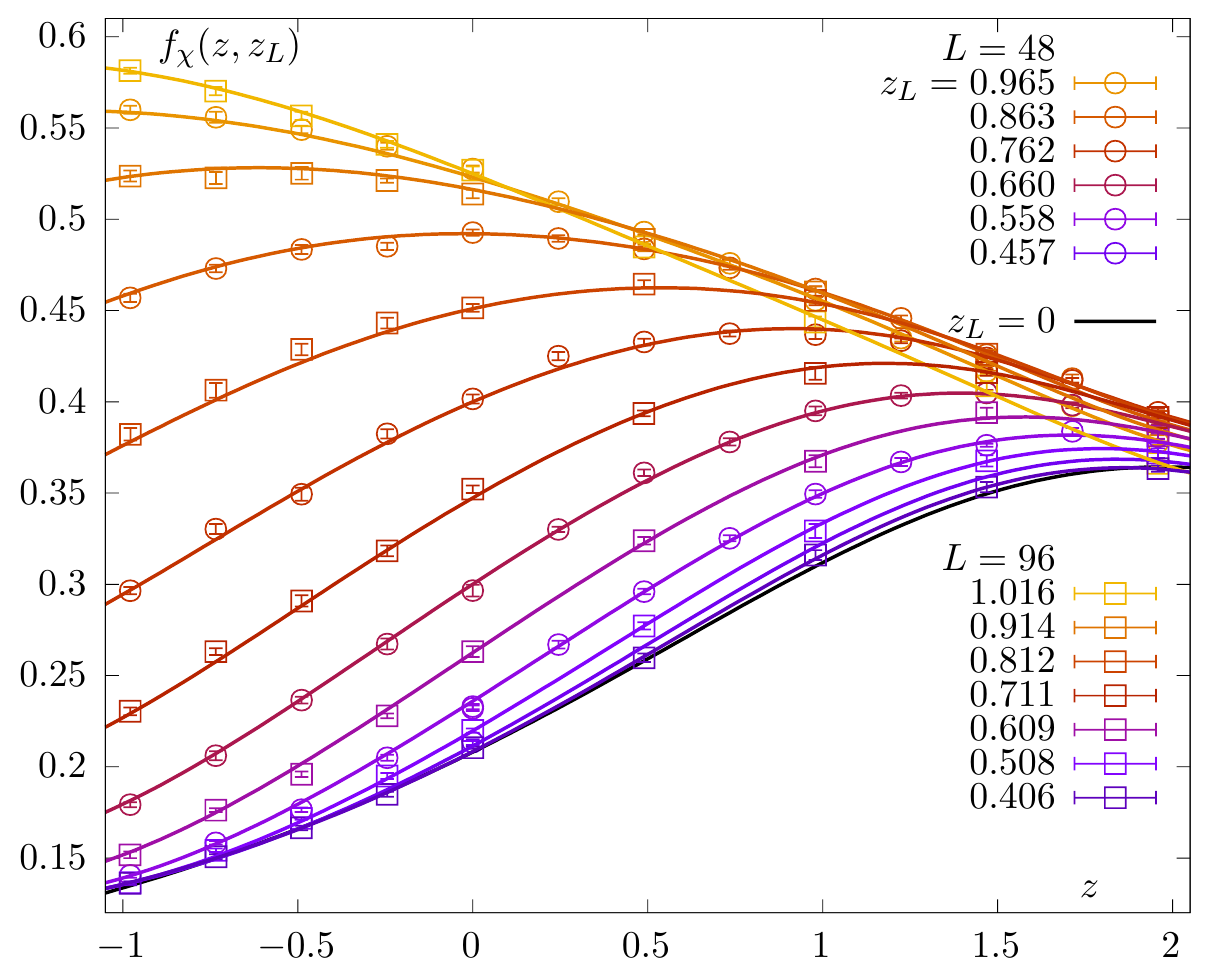}\hspace{0.7cm}
        \includegraphics[width=0.43\linewidth]{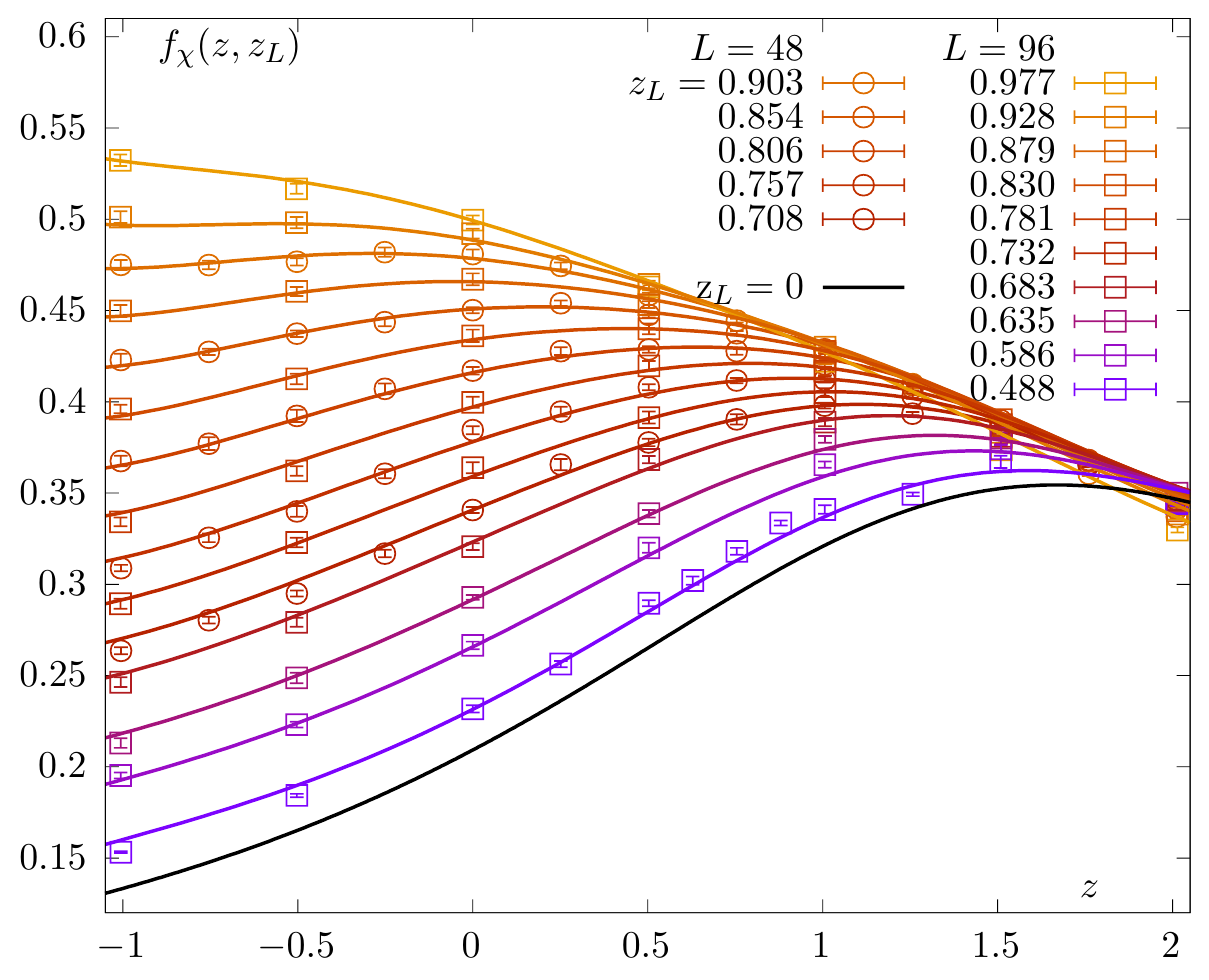}
    \caption{Fits to data for scaling functions of in the $Z(2)$ (left column) and $O(2)$ (right column)
    universality class. 
    All fits are joint fits to data for $f_G$, $f'_G$ and $f_\chi$, done 
    in the intervals $z\in [-1.0,2.0]$ and $z_L \in [0.4,1.0]$ on lattices of size $L=48$ (circle) and $96$ (squares). Also shown are the 
    infinite volume lines for
    $z_L=0$. Green crosses in the upper row mark the normalization condition $f_G(0,1)=0.7$.
}           \label{fig:Z2-O2-multifit}
\end{figure*}

\section{Finite size scaling functions}
\label{app:results}
We now want to determine corrections to the infinite volume scaling 
functions in the $3$-$d$, $Z(2)$ and $O(2)$ universality classes
arising in a finite volume at small external
field $H$. These corrections are universal
when taking the limit ($H\rightarrow 0$, $L\rightarrow \infty$) at fixed $z_L$ as introduced in Eq.~\ref{zTzL}. 

In the limit of small $H$ and in the 
vicinity of $T_c$, we obtain the
scaling functions $(f_G(z,z_L), f'_G(z,z_L), f_\chi(z,z_L))$ from 
the order parameter $M$ and the two susceptibilities $\chi_h$ and $\chi_t$
using Eqs.~\ref{fGzzL}-\ref{fchizzL}.

We focus here on the region in the 
vicinity of $T_c$ and the pseudo-critical temperatures, $T_{pc,h}$ and $T_{pc,t}$, 
determined from the maxima of the susceptibilities $\chi_h$ and $\chi_t$, respectively.
It is this region where 
correlation lengths are large and where
it is of particular importance to get control over finite-size effects in 
the determination of pseudo-critical
and critical temperatures in many models
belonging to the $Z(2)$ and $O(N)$ 
universality classes.
For this reason, we determine finite-size scaling functions with
parameter sets $(J,H)$ corresponding to the interval $z\in [-1:2]$.
A similar calculation has been 
performed previously for finite 
size scaling functions in the $O(4)$
universality class \cite{Engels:2014bra}.

In our analysis of finite-size effects, 
we use a polynomial ansatz for the scaling 
functions which has also been used
previously for calculations in the 3-$d$, $O(4)$ universality class
\cite{Engels:2014bra}, 
\begin{equation}
f_G(z,z_L) = f_G(z,0) + \sum^{n_u}_{n=0} \sum^{m_u}_{m=m_l} a_{nm} z^n z^m_L \; .
\label{FSSfG}
\end{equation}
For the infinite volume scaling function,
$f_G(z,0)\equiv f_G(z)$ we use the
parametrization determined in the previous section. Here, 
$(n_u,m_l,m_u)$ denote the lower and upper limits of the sum over 
the polynomial in powers of $z$ and $z_L$, respectively. We take 
the leading order finite-size correction to be inversely proportional 
to the volume $\mathcal{O}(1/L^3)$, \textit{i.e.}, $m_l=3$. The upper 
limits $n_u$ and $m_u$ are optimized in our fits, using the Bayesian 
information criterion. We fix $a_{0m_l}=0$ in both universality 
classes; additionally we constrain the fit parameters to $|a_{nm}|<10$.

From the ansatz used for $f_G(z,z_L)$, one also obtains the parametrization of
$f'_G(z,z_L)$, which controls the scaling
behavior of $\chi_t$,
\begin{equation}
f'_G(z,z_L) = f'_G(z,0) + \sum^{n_u}_{n=1} \sum^{m_u}_{m=m_l} n~ a_{nm} z^{n-1} z^m_L \; ,
\end{equation}
and  $f_\chi(z,z_L)$,  which controls the scaling
behavior of $\chi_h$,
\begin{eqnarray}
f_\chi ( z, z_L)  &=& 
f_\chi ( z,0)+ \label{FFS} \\
&& \sum^{n_u}_{n=0} \sum^{m_u}_{m=m_l} \left( \frac{1}{\delta} - \frac{n+m\nu}{\beta\delta} \right) a_{nm} z^n z^m_L\; .
\nonumber 
\end{eqnarray}
Using these polynomial ans\"atze we again perform 
joint fits to the MC data for the 
three scaling functions
($f_G, f'_G, f_\chi$) in the interval $z\in [-1:2]$ and for
$z_L\in [0.4:1.0]$. 
The data for $z_L<0.4$ have been excluded from these fits, as they have been used
already to determine the parameters of the infinite volume scaling functions,
as discussed in the previous section.

Results obtained for the finite-size 
scaling functions in the $3$-$d$, $Z(2)$ and $O(2)$ universality classes
for some fixed values of $z$ have been shown already in Fig.~\ref{fig:fixedz}. 
In Fig.~\ref{fig:Z2-O2-multifit}, we show
results for the scaling functions 
as function of $z$ for several fixed
values of $z_L$. The fit parameters
obtained with the polynomial fit ansatz, Eq.~\ref{FSSfG}, are given in Table~\ref{Z2_params} for 
the case of $Z(2)$ and in Table~\ref{O2_params} for the case of $O(2)$. These fits provide a good interpolation for our data in the range
$z_L\in [0.4:1.0]$. However, due to the 
large number of parameters involved, we 
cannot give significance to
individual parameters entering the
polynomial ansatz. We, therefore, quote
our fit result without assigning errors 
to the fit parameters.

As can be seen, the general $z_L$-dependence of
scaling functions $f_G(z,z_L)$ and 
$f_\chi(z,z_L)$ is similar in the $Z(2)$ and $O(2)$ universality
classes. However, it is apparent 
from the upper row in  Fig.~\ref{fig:Z2-O2-multifit} that
finite-size effects are larger in
the $O(2)$ case than for $Z(2)$.
In the latter case, results for $f_G(z,z_L)$ are indistinguishable 
from the infinite volume results 
already for $z_L< 0.6$, whereas 
in the $O(2)$ case at $z_L=0.6$, deviations
from the infinite volume values amount
to about 3\% at $z=-1$ and increase to 4\%
at $z=1$ (see also the discussion of Fig.~\ref{fig:fG_compared} in Appendix~\ref{app:H0L0}).
Furthermore, qualitative differences are evident in the $z_L$-dependence of the scaling function $f'_G(z,z_L)$. In the $Z(2)$ case, the approach to the infinite volume limit is non-monotonic for $z_L<0$. A pronounced peak shows up 
in the symmetry broken regime ($z\le 0$) at finite $z_l$,
and the asymptotic infinite volume limit is approached from above. In the 
case of the $O(2)$ universality class,
$f'_G(z,z_L)$ seems to approach the 
infinite volume limit result from below
for all $z$.

In the case of $f_\chi(z,z_L)$
the approach to the infinite volume 
limit is non-monotonic for $z$-values below the pseudo-critical scale, $z<z_p$.
As can be seen in Fig.~\ref{fig:fixedz},
this is the case in the $Z(2)$ as well
as in the $O(2)$ universality class.
This non-monotonic behavior is not that
prominently visible in  Fig.~\ref{fig:Z2-O2-multifit}, as it 
sets in only at rather large values 
of $z_L$, {\it i.e.} for $z_L > 1$. This 
regime is not covered in Fig.~\ref{fig:Z2-O2-multifit}.

\begin{figure}[t]
    \centering
        \includegraphics[width=0.85\linewidth]{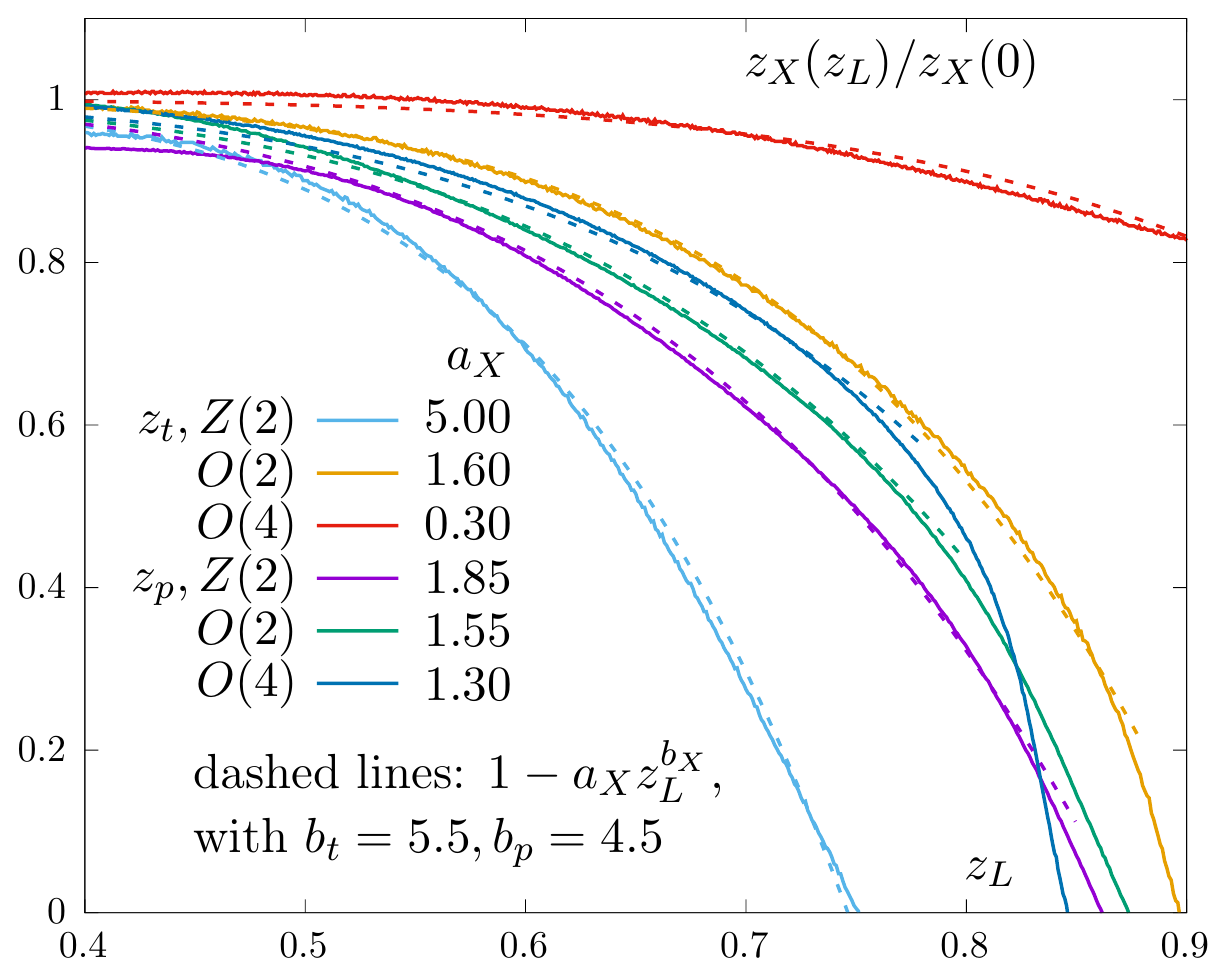}
    \caption{Finite-size dependence of the location of maxima in the scaling functions $f_\chi(z,z_L)$ and $-f'_G(z,z_L)$. Shown are 
    the universal functions $z_p(z_L)$ and $z_t(z_L)$ for the $3$-$d$, $Z(2)$ and $O(N)$ universality classes. Dashed lines show simple polynomial approximations (Eq.~\ref{ztzpapprox}) with parameters as given in the figure.}
    \label{fig:maxima}
\end{figure}

While the finite-size effects seen in the 
scaling functions are generally larger in the $O(N)$ than in the $Z(2)$ 
universality class, this is not the case for the location of maxima in the 
scaling functions $-f'_G(z,z_L)$ and $f_\chi(z,z_L)$. These maxima are 
controlled by universal functions $z_t(z_L)$ and $z_p(z_L)$, respectively. 
We determined them from the polynomial obtained from the finite-size scaling 
fits for $Z(2)$ and $O(2)$. In the case of $O(4)$, we have used the finite-size 
fit given in \cite{Engels:2014bra}. Results are shown in Fig.~\ref{fig:maxima}.
It is clearly seen that the finite-size
dependence of the maxima in $\chi_h$
is stronger than that of $\chi_t$ in the 
$O(N)$ universality classes and vice versa in the $Z(2)$ case. Moreover, the finite
size dependence of $z_t$ and $z_p$ is
stronger in the $Z(2)$ universality class than in the $O(N)$ cases. 
Over a wide range of $z_L$-values,
the deviations from the infinite
volume limit result are described 
well with ansatz 
\begin{equation}
    z_X(z_L) =z_X(0) \left( 1-a_X z_L^{b_X}\right) \; ,\; X=p,\ t \; ,
    \label{ztzpapprox}
\end{equation}
with $b_p\simeq 4.5$ and $b_t\simeq 5.5$ as shown in Fig.~\ref{fig:maxima}.

\begin{figure*}[htb]
        \centering
        \begin{tabular}{|c|ccccc|} \hline
            $a_{nm}$ & $n=0$ & $n=1$ & $n=2$ & $n=3$ & $n=4$ \\ \hline
            $m=3$ & 0          & -0.948309 & 0.717317 & -0.162262 & 0.077211 \\
            $m=4$ & -1.626176 & 5.613893 & -3.665684 & 0.705625 & -0.389709 \\
            $m=5$ & 7.182912 & -9.594472 & 4.926683 & -1.015233 & 0.782710 \\
            $m=6$ & -7.151294 & 0.244453 & 0.803864 & 0.708351 & -0.624599 \\
            $m=7$ & -6.583527 & 7.559574 & -3.189855 & 0.340024 & -0.444347 \\
            $m=8$ & 7.641469 & 2.794132 & -1.881236 & -1.199165 & 0.834052 \\
            $m=9$ & 7.510637 & -4.712291 & 1.481824 & -0.473266 & 0.548429 \\
            $m=10$ & -9.932439 & -4.460071 & 2.132431 & 2.098657 & -1.309965 \\
            $m=11$ & 2.658208 & 3.544670 & -1.293838 & -1.003939 & 0.524142 \\ \hline
        \end{tabular}
        \captionof{table}{Parameters of the polynomial fit ansatz for the $Z(2)$ finite-size scaling functions $(f_G, f'_G, f_\chi)(z,z_L)$ with $n_u=4$, $m_l=3$, and $m_u=11$. The fit was restricted to $z\in [-1:2]$ and $z_L \in [0.4,1.0]$.}
        \label{Z2_params}
\end{figure*}

\begin{figure*}[htb]
        \centering
        \begin{tabular}{|c|cccccc|} \hline
            $a_{nm}$ & $n=0$ & $n=1$ & $n=2$ & $n=3$ & $n=4$ & $n=5$ \\ \hline
            $m=3$ & 0         & -0.740936 & 0.198298 & 0.020480 & -0.219484 & 0.096802 \\
            $m=4$ & -0.735344 & 4.506235 & -0.871942 & -0.015341 & 1.377204 & -0.676093 \\
            $m=5$ & 4.031332 & -9.950240 & 0.340937 & -0.456753 & -3.161604 & 1.800423 \\
            $m=6$ & -9.769988 & 9.634183 & 2.604256 & 1.621876 & 3.397707 & -2.365422 \\
            $m=7$ & 8.841449 & -3.490053 & -3.497933 & -1.908686 & -1.770614 & 1.560082 \\
            $m=8$ & -2.670471 & 0.113312 & 1.259727 & 0.735997 & 0.371330 & -0.414748 \\ \hline
        \end{tabular}
        \captionof{table}{
        Parameters of the polynomial fit ansatz for the $O(2)$ finite-size scaling function $(f_G, f'_G, f_\chi)(z,z_L)$ with $n_u=5$, $m_l=3$, and $m_u=8$. The fit was restricted to $z\in [-1:2]$ and $z_L \in [0.4,1.0]$.
        }
        \label{O2_params}
\end{figure*}

\section{Conclusions}

We determined the infinite volume scaling functions in the $3$-$d$, $Z(2)$, $O(2)$, and $O(4)$ universality classes using a 2 or 3
parameter parametrization based on the analytic Widom-Griffiths scaling
form.
We find good agreement of the $O(4)$
parametrization with an earlier
parametrization that used ${\cal O}(10)$ parameters \cite{Engels:2011km}. In the $Z(2)$ 
case, we find excellent agreement between our parametrization based 
on Monte Carlo results and the analytic result obtained from a perturbative, field theoretic approach \cite{Zinn-Justin:1999opn}. The largest differences between our Monte Carlo results and analytic calculations \cite{Campostrini:2000iw} we find, in particular, for the scaling function $f'_G(z)$, which controls the scaling behavior of mixed susceptibilities.

We determined the finite-size dependence of the scaling functions and showed that
qualitative differences between the $Z(2)$ and $O(N)$ cases show up most prominently in the scaling function $f'_G(z,z_L)$ which controls pseudo-critical and critical behavior of 
the mixed susceptibilities. 
We could show that the location of the pseudo-critical temperature, corresponding to $z_t$, is less affected by finite-size 
effects than the pseudo-critical 
temperature determined by the maximum of
the order parameter susceptibility ($\chi_h$) at $z_p$. 
This difference is particularly striking in the $O(4)$ universality class. The comparison of the finite-size dependence of the scaling functions among different universality classes has been possible with our proposed normalization condition for the non-universal scale parameter $L_0$. 

We furthermore find, $z_p/z_t\simeq 2$, {\it i.e.} at non-zero values of 
the symmetry breaking parameter $H$ deviations of the pseudo-critical 
temperature $T_{pc,t}$ from the phase transition
temperature $T_c$ are about a factor of 2 smaller than that of $T_{pc,h}$. 
All data presented in the figures of this paper can be found in Ref.~\cite{karsch2023dataset}.

\section*{Acknowledgments}
\label{sec:acknowledge}
This work was supported in part by the Deutsche Forschungsgemeinschaft (DFG) through the grant 315477589-TRR 211
and also supported in part by the National Science and Technology Council, the Ministry of Education (Higher Education Sprout Project NTU-112L104022), and the National Center for Theoretical Sciences of Taiwan.
All calculations have been performed on the Bielefeld University GPU cluster and the GPU cluster of the Paderborn Center for Parallel Computing 
($PC^2$).
We thank J\"urgen Engels for making his Monte Carlo simulation code available to us, which has been used for all calculations presented here. We also thank Anirban Lahiri for many helpful discussions and his valuable contributions in the early phase of this research project.

\appendix
\section{Determination of \boldmath$H_0$ and $L_0$}
\label{app:H0L0}
\begin{figure*}[t]
    \centering
       \includegraphics[width=0.45\linewidth]{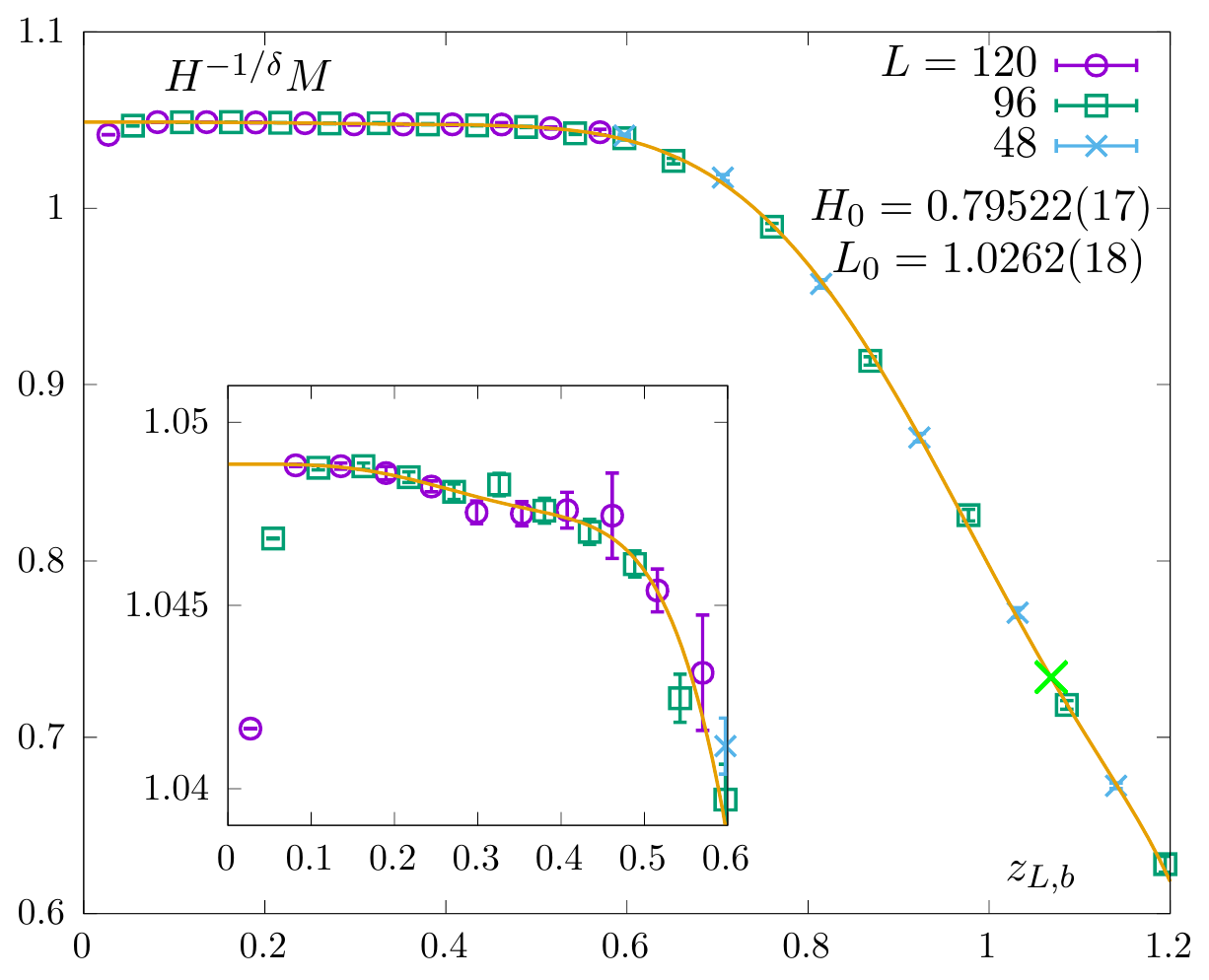}
    \includegraphics[width=0.45\linewidth]{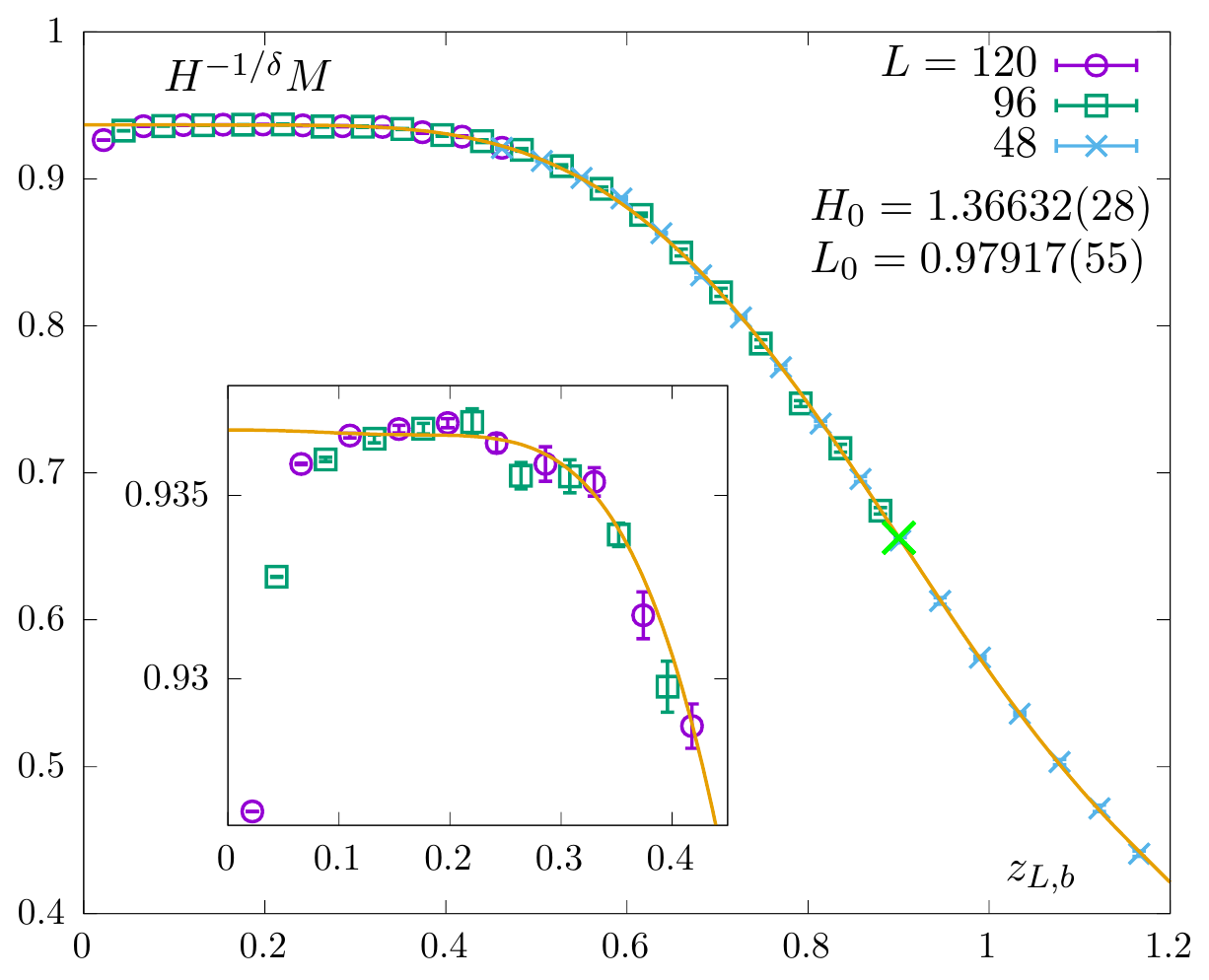}
    \caption{
    The rescaled order parameter $H^{-1/\delta} M$
    versus the bare finite-size scaling variable $z_{L,b}$
    at $T_c$. The left hand figure shows results for the $3$-$d$,
    $Z(2)$ model with $\lambda=1.1$, and the right hand figure is for the $O(2)$ model with $\lambda=2.1$.
    The inset shows the region of small $z_{L,b}$ that is used for the
    determination of the non-universal scale parameter $H_0$. The green cross marks the value of $z_{L,b}$ that determines the scale parameter $L_0$
    using the normalization condition given in Eq.~\ref{L01}. 
    }
    \label{fig:O2_H0}
\end{figure*}

In order to extract scaling functions from numerical simulations of the $3$-$d$, $Z(2)$ and $O(2)$ 
model using the Hamiltonian given
in Eq.~\ref{Hamiltonian}, we need
to determine the non-universal scales, $(t_0, H_0, L_0)$. 
In this appendix, we discuss the determination of $(H_0, L_0)$
using the finite-size dependence of the order parameter at $T_c$.

The critical temperature $T_c$ has been determined with great 
precision for the improved $Z(2)$  \cite{Hasenbusch:1999mw} and 
$O(2)$ \cite{Campostrini:2000iw} models, respectively. 
For the $Z(2)$ model, also the scale $H_0$ has been determined previously
\cite{Engels:2002fi} on similar size lattices as used in this 
study but using infinite volume scaling ans\"atze and lower statistics.

For the determination of $H_0$ we make use of the normalization conditions 
for the order parameter or, equivalently, the scaling function $f_G(z,0)$ as introduced in Eq.~\ref{fGconditions}.
The scale $L_0$ is obtained using the normalization condition
for the finite-size scaling function $f_G(0,z_L)$ introduced in Eq.~\ref{L01}.

For our determination of the scale parameters, we introduce the (bare) scaling 
variables $z_b$ and $z_{L,b}$ through $z=z_0 z_b$ and $z_L =z_{0,L} z_{L,b}$, with 
\begin{eqnarray}
    z_b &=& \frac{T-T_c}{T_c} H^{-1/\beta\delta} \;\; ,
    \label{zb} \\
    z_{L,b} &=&\frac{1}{LH^{\nu/\beta\delta}}\; ,
 \label{zLb}
 \end{eqnarray}
 and
 \begin{eqnarray}
    z_0 = H_0^{1/\beta\delta}/t_0\;\; ,\;\; 
    z_{0,L}= L_0 H_0^{\nu/\beta\delta}\; .
    \label{z0}
\end{eqnarray}

To determine $H_0$, using  Eq.~\ref{fGconditions}, we performed
dedicated calculations at $T_c$ 
on lattices of size $L=48,\ 96$ 
and $120$ and for several values of $z_L$. 
The statistics collected for
each parameter set $(J_c,L)$ is given in Tab.~\ref{tab:statistics_H0}.
We calculate the order parameter $M(T,H,L)$
in the limit $(H\rightarrow 0,L\rightarrow \infty)$ for several
values of fixed $z_{L,b}$ and 
then take the limit $z_{L,b}\rightarrow 0$ at $T\equiv T_c$,
\begin{equation}
H_0^{-1/\delta} =
\lim_{z_{L,b}\rightarrow 0}
    \lim_{H\rightarrow 0}
\left(H^{-1/\delta} M(T_c,H,1/z_{L,b} H^{\nu/\beta\delta})\right) \; .
\label{H0-det}
\end{equation}
Results from this calculation 
are shown in Fig.~\ref{fig:O2_H0}. The intercept at $z_{L,b}=0$ yields $H_0^{-1/\delta}$.
Also shown in the figure are results from polynomial fits,
\begin{equation}
    \tilde{f}_G(z_{L,b}) = H_0^{-1/\delta} + \sum^{m_u}_{m=m_l} b_m z^m_{L,b}
\; ,
\label{H0-fit}
\end{equation}
to the right hand side of Eq.~\ref{H0-det} in different intervals  $z_{L,b}\in[0.1:z_{L,b,\text{max}}]$, with $z_{L,b,\text{max}}\in\{1.1,1.2,1.3\}$.
$H_0$ and $b_m$ are then determined by bootstrapping fits with different $z_{L,b,\text{max}}$.
The lower and upper limits $m_l$ and $m_u$ are chosen differently from their finite-size counterparts: We use $m_l=4$ and $m_u=9$ for $Z(2)$, while $m_l=3$ and $m_u=7$ are used for $O(2)$. 
This determines $H_0$. Using Eq.~\ref{L01}, we then obtain $L_0$ from the 
value $z_{L,b}$, which gives $H_0^{1/\delta} \tilde{f}_G(z_{L,b}) =0.7$. 
Using the fit results for $\tilde{f}_G(z_{L,b})$, we then obtain the normalization constants $(H_0,L_0)$ for the $Z(2)$ and
$O(2)$ model, which are given in
Table~\ref{tab:parameter}.

The result obtained for $H_0$ for the $Z(2)$ model from our finite-size 
scaling fit, is about 2\% smaller than the value $H_0=0.8150(56)$ obtained 
in \cite{Engels:2002fi} from a fit of the order parameter $M$ at $T_c$,
using the infinite volume scaling ansatz for $M$. 

\begin{figure}[t]
\includegraphics[width=0.8\linewidth]{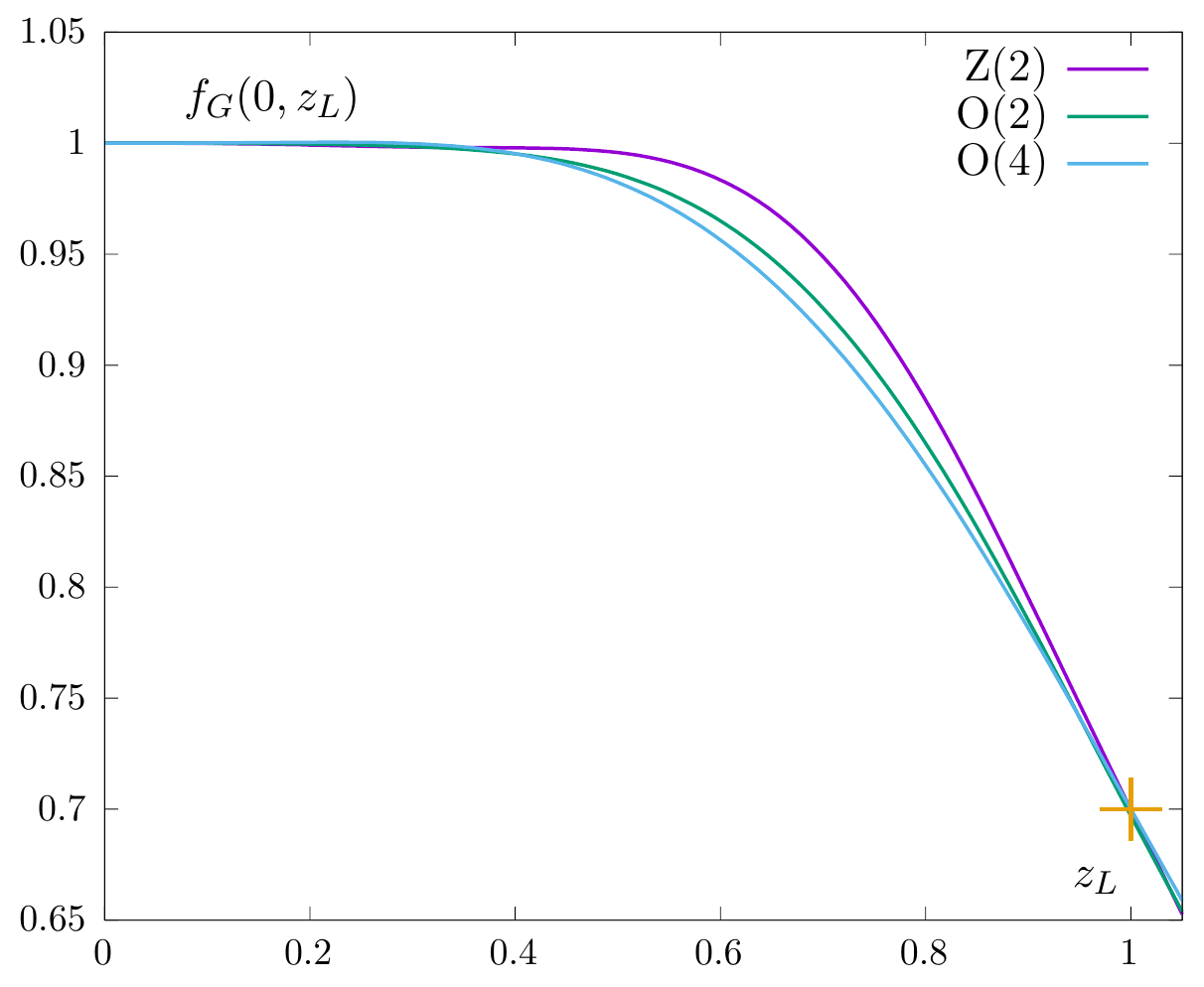}
\caption{Comparison of the scaling function $f_G(0,z_L)$
in different $3$-$d$ universality classes as a function of $z_L$. The data 
for the scaling function in the $O(4)$ universality class are taken from 
\cite{Engels:2014bra}. For this purpose, the scaling variable $z_L$ has 
been rescaled using $L_0=0.7686$ to be consistent with the normalization 
condition, Eq.~\ref{L01}, used for the $Z(2)$ and $O(2)$ universality classes.}
\label{fig:fG_compared}
\end{figure}

Using the scale parameters $H_0$ and $L_0$, we obtain the scaling function 
$f_G(z,z_L)$ at $z=0$ as a function of $z_L$. A comparison of results obtained 
in different universality classes is shown in Fig.~\ref{fig:fG_compared}.
This suggests that the finite-size dependence of the order parameter is larger 
in the $O(N)$ universality classes than in the $Z(2)$ universality class.

\section{Parametrization of \boldmath$Z(2)$ and $O(N)$ scaling functions}
\label{app:Widom}
We give here results for the two sub-leading expansion coefficients, $d_1^-$ and $d_2^-$,  appearing in the large, negative $z$ expansion of the infinite volume scaling functions
$f_G(z)$ in the $3$-$d$, $Z(2)$ and $O(N)$ universality classes (\textit{cf.} Eq.~\ref{fGzb}).
We present explicit expressions in terms of the parameters appearing 
in the definition of the function 
$h(\theta)$ given in Eq.~\ref{hthetaZ2O2} of scaling functions
written in the Widom-Griffiths form \cite{Widom,PhysRev.158.176,PhysRevLett.22.606,PhysRevLett.23.1098}.

The coefficients in the asymptotic expansion for the 
$Z(2)$ scaling function are
\begin{eqnarray}
d_1^-&=&
-\theta_0^{\delta -1}\frac{(1+(2 \beta -1)\theta_0^2) }{(\theta_0^2-1) } \frac{h(1)}{h'(\theta_0)}
 \; ,  
 \end{eqnarray}
 \begin{eqnarray}
	d_2^-&=&
	   -\frac{ \theta_0^{2 \delta -1} h(1)^2}{2 \left(\theta_0^2-1\right)^2 h'(\theta_0)^3} \nonumber \\
	   &&\Biggl( 2\beta\theta_0    h'(\theta_0)
    \left(2\delta  \left((2 \beta -1) \theta_0^2+1\right)\right. \nonumber \\
    &&\left.  -(2 \beta 
    -1)\theta_0^2-3\right)
    \nonumber \\
    && -h^{(2)}(\theta_0) \left(\theta_0^2-1\right)
    \left((2 \beta -1) \theta_0^2+1\right)\Biggr) \; ,
\end{eqnarray}
and the corresponding expansion coefficients in the $O(N)$ case are
    \begin{eqnarray}
    d_1^- &=&
    \theta_0^{\delta/2-1}\frac{ \left(1+(2 \beta -1) \theta_0^2\right)}{
    \left(\theta_0^2-1\right)}
    \sqrt{\frac{2 h(1)}{
    h^{(2)}(\theta_0)}}
    \\ 
    d_2^- &=&
-\frac{\theta_0^{\delta -1} h(1)}{3
    \left(\theta_0^2-1\right)^2 h^{(2)}(\theta_0)^2}  \nonumber \\ &&\Biggl(6 \beta  \theta_0 h^{(2)}(\theta_0) \left(\delta 
    \left((2 \beta -1) \theta_0^2+1\right)\right. \nonumber \\
    &&\left. -(2 \beta -1) \theta_0^2-3\right)
     \nonumber \\
    && - h^{(3)}(\theta_0)\left(\theta_0^2-1\right) \left((2 \beta -1)
    \theta_0^2+1\right)\Biggr) \; .
\end{eqnarray}
It should be noted that $\theta_0$ is an independent parameter in the parametric representation of $O(N)$ universality class while it is a function of parameters $h_3$ and $h_5$ in the $Z(2)$ case.

%

\end{document}